\documentclass[conference,10pt]{IEEEtran}
\usepackage{times}
\usepackage{psfig}
\usepackage{graphicx}
\usepackage{subfigure}

\begin{document}

\title{Stationary and Mobile Target Detection using Mobile Wireless Sensor Networks}

\author{Ev\c{s}en Yanmaz$^1$ and Hasan Guclu$^2$\\
\IEEEauthorblockA{$^1$Institute of Networked and Embedded Systems, Mobile Systems Group,
University of Klagenfurt, Austria\\
$^2$ School of Mathematical Sciences, Rochester Institute of Technology, Rochester, NY 14623\\
Email: evsen.yanmaz@uni-klu.ac.at}}
 

\maketitle
\thispagestyle{empty}

\begin{abstract}
In this work, we study the target detection and tracking problem in mobile sensor networks, where the performance metrics of interest are probability of detection and tracking coverage, when the target can be stationary or mobile and its duration is finite. We propose a physical coverage-based mobility model, where the mobile sensor nodes move such that the overlap between the covered areas by different mobile nodes is small. It is shown that for stationary target scenario the proposed mobility model can achieve a desired detection probability with a significantly lower number of mobile nodes especially when the detection requirements are highly stringent. Similarly, when the target is mobile the coverage-based mobility model produces a consistently higher detection probability compared to other models under investigation.
\end{abstract}

\section{Introduction}

Target or event detection/tracking has been one of the main applications of wireless sensor networks. While most of the previous work has been on static sensor networks, recently it has been shown that the coverage of a sensor network can be improved via mobility \cite{Liu-Towsley2}. In case of monitoring geographically inaccessible or dangerous areas, mobile robots equipped with sensors can be deployed for effective coverage. Moreover, if the target (event) to be detected by the sensor network is of time-critical nature, the coverage of the network should be sufficiently high to be able to respond to the detected event in a timely manner; such as wildfire monitoring or liveliness detection under rubble in case of an earthquake, where the emergency personnel work against the clock. 

To this end, in this paper, we propose a coverage-based cooperative mobility model for wireless sensor networks to detect and track (monitor) targets without prior knowledge of the physical topology and by using only local topology information. While determining the mobility path, no assumption is made on the application the sensor network is deployed for. Empirical studies are conducted to test the performance of the proposed model, where a stationary target (such as a live body under rubble) or a mobile target (such as animals monitored in their habitat) is assumed to occur at a random location in the geographical area to be monitored and target detection probability and tracking (monitoring) efficiency performance of the proposed model is compared with legacy mobility models such as random walk, random direction, etc. It is shown that the coverage-based mobility model consistently results in a better performance than the other mobility models. In addition, we also provide a brief analysis to determine the minimum number of nodes required to achieve a certain target coverage, for the stationary target scenario. Results show that while for small detection probabilities all mobility models perform similarly, for higher desired detection probabilities coverage-based mobility model significantly outperforms the rest in terms of the number of required mobile nodes. 

The remainder of the paper is organized as follows. In Section II background on mobility models and coverage problem in wireless sensor networks is summarized. The proposed coverage-based mobility model is presented in Section III. A brief detection analysis is provided in Section IV. Results are given in Section V and the paper is concluded in Section VI.

\section{Background}

\subsection{Mobility models}
The system under investigation is a wireless sensor network that consists of only mobile nodes with the same transmission range. The system parameters are summarized in Table~\ref{SysPar}. There are several mobility models that consider independent or dependent movement among mobile nodes \cite{survey1}. In this paper, the following well-known mobility models are considered: 

$\bullet$ {\em Random Walk}: A mobile node picks a random speed and direction from pre-defined uniform distributions either at fixed time intervals or after a certain fixed distance is traveled. The current speed and direction of the mobile node do not depend on the previous speeds and directions.


$\bullet$ {\em Random Direction}: A random direction drawn from a uniform distribution is assigned to a mobile node and the mobile node travels in that direction till it reaches the boundary of the simulation area. Once it reaches the boundary, it pauses there for a fixed amount of time, then moves along the new randomly selected direction. In this paper, for fair comparison, we assume that the pause time is zero.

$\bullet$ {\em Parallel-path}: A mobile node picks a random speed and sweeps the geographical area from border to border following a direction parallel to the boundary line.

The models explained above are memoryless, i.e., the current directions are independent of the past directions and the mobile nodes decide their movements independently from each other. There are several other mobility models that take into account the dependence on the mobility pattern of other nodes in the network \cite{survey1}, social relationships of the mobile nodes \cite{CommBased}, or topographical information \cite{ObsMob}, etc. In this work, initially, the models explained above are studied in addition to the proposed mobility model in Section III. Note that while in the above models (and the proposed model presented in the next section) the speeds are drawn from a probability distribution, in the Monte Carlo simulations conducted in this paper we assume that the mobile nodes have a fixed speed.

\subsection{Coverage in Wireless Sensor Networks}
Coverage problem in wireless sensor networks is of great importance and has been investigated by several researchers. In static wireless sensor networks, in general, coverage problem is treated as a node-activation and scheduling problem \cite{IntCovCon1}-\cite{Liu-Towsley}. More specifically, algorithms are proposed to determine which sensor nodes should be active such that an optimization criterion is satisfied. The criterion can for instance be achieving a certain event detection probability, or covering each point in the area by at least $k$ sensors, etc. In addition, there are also studies that take into account not only the event (or network) coverage, but the connectivity of the wireless sensor network as well \cite{IntCovCon1}. While deciding which sensor nodes should be active at a given point in time, coverage and connectivity requirements are met.

Recently, mobile sensor networks have been under investigation and it has been shown that mobility, while complicates the design of higher layer algorithms, also can improve the network, for instance, in terms of capacity, coverage, etc. \cite{MobCap}-\cite{Liu-Towsley2} Optimum mobility patterns for certain applications are proposed, such as mobile target tracking, chemical detection, etc. using both ground and aerial vehicles. Mobile robots with swarming capability operate cooperatively and aim to achieve a global goal have also been considered \cite{ConsCov}-\cite{CoopRelay}. 

In robotics, several mobility models have been developed. In many of these models the robots which are too close repel each other to avoid collisions but to maintain communication they attract each other when they are separated more than certain distance. Gas expansion model \cite{payton}, for example, mimics the way gas particles are spread to vacuum when they are allowed to expand. This model, again, uses the attraction and repulsion forces between the robots to maximize the dispersion while maintaining the communication. Similar models have also been proposed by using an artificial force or potential fields for the robots to cooperatively move \cite{spears, pereira}. However, the focus in these studies is maximizing the spread not the coverage, and they are based on the assumption that the robots have high computing capacities.

In this work, without assuming any predetermined application, we propose a collaborative mobility model based on only local information to improve area coverage in wireless sensor networks with low computation-capability nodes.

\begin{table}
\caption{System Parameters}
\label{SysPar}
\centering
\begin{tabular}{|l|l|} \hline
Parameter & Definition \\ \hline
$N_m$ & Number of mobile nodes \\ \hline
$r$ & Transmission range \\ \hline
$a$ & Square simulation area length \\ \hline
$\rho$ & Spatial node density \\ \hline
$P_d$ & Event (target) detection probability \\ \hline
$t_d$ & Event (target) duration \\ \hline
\end{tabular}
\end{table}

\section{Coverage-based Mobility Model}
In this section, we propose a mobility model that makes use of the local physical topology information. The objective is to improve coverage of a geographical area for a given mobile sensor network. While designing the model, the application is not predetermined. More specifically, only the physical information is used to determine the mobility path. The performance of the algorithm will be tested for an event detection/monitoring application later in the paper. 

We assume that there is no prior knowledge of the location of the mobile nodes. Since the objective is to improve coverage, it is desirable to reduce the overlap between the covered areas by different mobile nodes and use the limited number of mobile nodes efficiently (specifically, if the mobile sensor network will be used for a time-critical application.).  

\begin{figure}[!htb]
\centerline{\psfig{figure=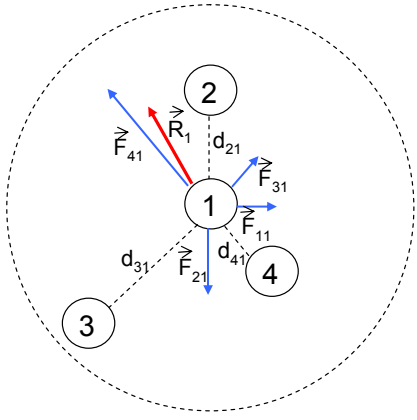,width=0.3\textwidth}}
\caption{Illustration of forces on mobile node 1, where the dashed circle is the transmission range of the node and mobile node 1 is moving toward right.}
\label{CBForce}
\end{figure}

In the model, the speed of the mobile nodes is a uniform random variable in $[0,V_m]$ and the direction is chosen in fixed time intervals according to the local topology at the time of the decision. More specifically, we assume that there is a force between mobile nodes that causes them to {\em repel} each other. The magnitude of the force that each node applies to others is inversely proportional to the distance between the nodes, i.e., the closer the nodes get the stronger they {\em push} each other. We also assume that the mobile node knows its current direction and a force with a magnitude inversely proportional to the node's transmission range (i.e., $r$) is applied to it in the direction of movement to avoid retracing the already covered areas by the mobile node. At the time of direction change, each mobile node computes the {\em resultant} force vector acting on them by themselves and their neighbors (i.e., the mobile nodes within their transmission range) and move in the direction of the resultant vector. The forces at the time of decision are illustrated in Fig.~\ref{CBForce}, where mobile node 1 is moving toward right in the previous step. While the distance between the mobile nodes can easily be determined if the nodes are equipped with GPS, due to cost limitations it might be more feasible to use the received signal strength jointly with the direction of signal arrival to estimate the distance. Further work is necessary to determine a power and cost efficient method to estimate the distances between mobile nodes.

Observe from Fig.~\ref{CBForce} that the resultant force on node $i$, $\vec{R}_{i} = \sum_j{\vec{F}_{ji}}$, where $\vec{F}_{ii} \parallel \vec{V}_i$ with $\left|\vec{F}_{ii}\right| = \frac{1}{r}$ and $\left|\vec{F}_{ji}\right| = \frac{1}{d_{ji}}$ when $j \neq i$, where $\vec{V}_i$ is the velocity vector of mobile node $i$, $r$ is the transmission range of each mobile node, and $d_{ij}$ is the distance between nodes $i$ and $j$. The direction of $\vec{F}_{ji}$ is parallel to the line drawn from node $j$ to node $i$. 
Mobile node $i$ will move in the direction of $\vec{R}_{i}$ with a speed chosen from the range $[0,V_m]$ for a fixed time duration (i.e., a step length). Same algorithm is run for all the mobile nodes and the directions are updated accordingly. If, at the time of direction change, a mobile mode does not have any neighbors, the direction is not changed. Note that the step length is a design parameter and depends on the system parameters such as $N_m$ and $r$ among others. Optimum step length is currently under investigation. 

\section{Detection Analysis}

In this section, we provide an approximation to the event coverage (detection) probability by the mobile nodes within a given time duration $t$. Note that the event detection probability can be determined from the percentage area that is covered over time $t$. Assume that the transmission range of the mobile nodes is $r$ and their coverage area is of disc shape, i.e., area covered by each node is $\pi r^2$. 

First, let's assume that the nodes are static. Given that the total area to be covered is $A$, the number of sensor nodes is $N$, and the initial locations of the sensor nodes are uniformly random, the number of sensor nodes that cover a given point $i$ in the observed area has a Poisson distribution with parameter $\rho \pi r^2$, where $\rho = N/A$ \cite{Liu-Towsley2}:
\begin{eqnarray}
Prob\{{\mbox{i is covered by k nodes}}\} = \frac{e^{-\rho \pi r^2}(\rho \pi r^2)^k}{k!}
\label{Poi}
\end{eqnarray}

Therefore, the probability that a point $i$ is not covered by any nodes is given by: 
\begin{eqnarray}
Prob\{\mbox{i is not covered (k = 0)}\} = e^{-\rho \pi r^2}
\label{Poi2} 
\end{eqnarray}

Integrating Eq.~(\ref{Poi2}) for all points $i$ in the area $A$ and normalizing with respect to the total area to be covered, the percentage covered area (i.e., event detection probability) by one or more static sensor nodes can be shown to be:

\begin{eqnarray}
\nonumber
P_{c_s} &=& 1-e^{-\rho \pi r^2} \\
&=& 1-e^{-N \pi r^2/A}
\label{Pc_anl}
\end{eqnarray}

From Eq.~(\ref{Pc_anl}), the minimum number of static sensor nodes necessary to cover a geographical area (i.e., detect an event) with probability $P_d$, where $0<P_d<1$ is given as:
\begin{eqnarray}
N^{min}_s = \frac{-A\ln(1-P_d)}{\pi r^2}
\label{Nmin_anl}
\end{eqnarray}

\begin{figure}[!htb]
\centerline{\psfig{figure=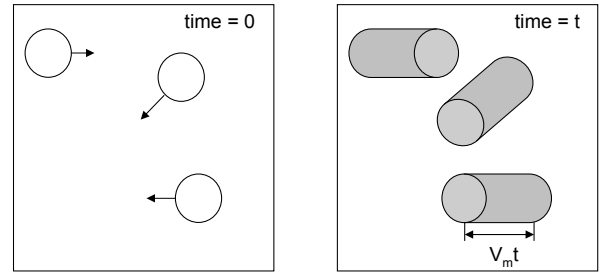,width=0.45\textwidth}}
\caption{Coverage illustration of mobile sensor nodes during time duration $t$, when the speed of the mobile nodes is $V_m$.}
\label{CovIll}
\end{figure}

Next, we will find the percentage covered area by mobile nodes during $t$. As shown in Fig.~\ref{CovIll}, the covered area by a mobile node in $[0,t]$ can be represented as the union of discs. Assume that the speed and the direction of the mobile nodes are independent and identically distributed with probability distribution functions $f_V(v)$ and $f_\Theta(\theta)$, respectively, where $v\in [0,V_m]$ and $\theta \in [0,2\pi]$. Observe from Fig.~\ref{CovIll} that the effective area covered by each sensor node increases and the average covered area changes from $\pi r^2$ to $\pi r^2+2rE[V]t$ at time $t$, when the nodes are mobile, where $E[V]$ is the expected value of the sensor speed. Since the mobile nodes move independently from each other, the distribution of mobile nodes at any time instant still has a Poisson distribution \cite{Liu-Towsley2}. Therefore, similar to the static case, the percentage of covered area by at least one mobile sensor node in $[0,t]$ can be shown to be:

\begin{eqnarray}
\nonumber
P_{c_m} &=& 1-e^{-\rho (\pi r^2+2rE[V]t)} \\
&=& 1-e^{-N (\pi r^2+2rE[V]t)/A}
\label{Pcm_anl}
\end{eqnarray}

From Eq.~(\ref{Pcm_anl}), the minimum number of mobile sensor nodes necessary to cover a geographical area (i.e., to detect an event) with probability $P_d$ within a time duration $t$, where $0<P_d<1$ is given as:
\begin{eqnarray}
N^{min}_{m} = \frac{-A\ln(1-P_d)}{\pi r^2+2rE[V]t}
\label{Nmmin_anl}
\end{eqnarray}

On the other hand, if a mobility model exists such that the coverage of mobile nodes do not overlap at any time, detection can clearly be achieved with a smaller number of nodes. For a {\em non-overlapping} mobility model, the number of required nodes can be calculated to be:
\begin{eqnarray}
N_{nooverlap} = \left\lceil{\frac{A}{\pi r^2+2rE[V]t_d}}\right\rceil
\end{eqnarray}
However, since the node distribution of the random mobility models used in this paper exhibits a uniformly random distribution, we use the equations for random network in the next section. 

\begin{figure}[!htb]
\centering
\subfigure[]
{\psfig{figure=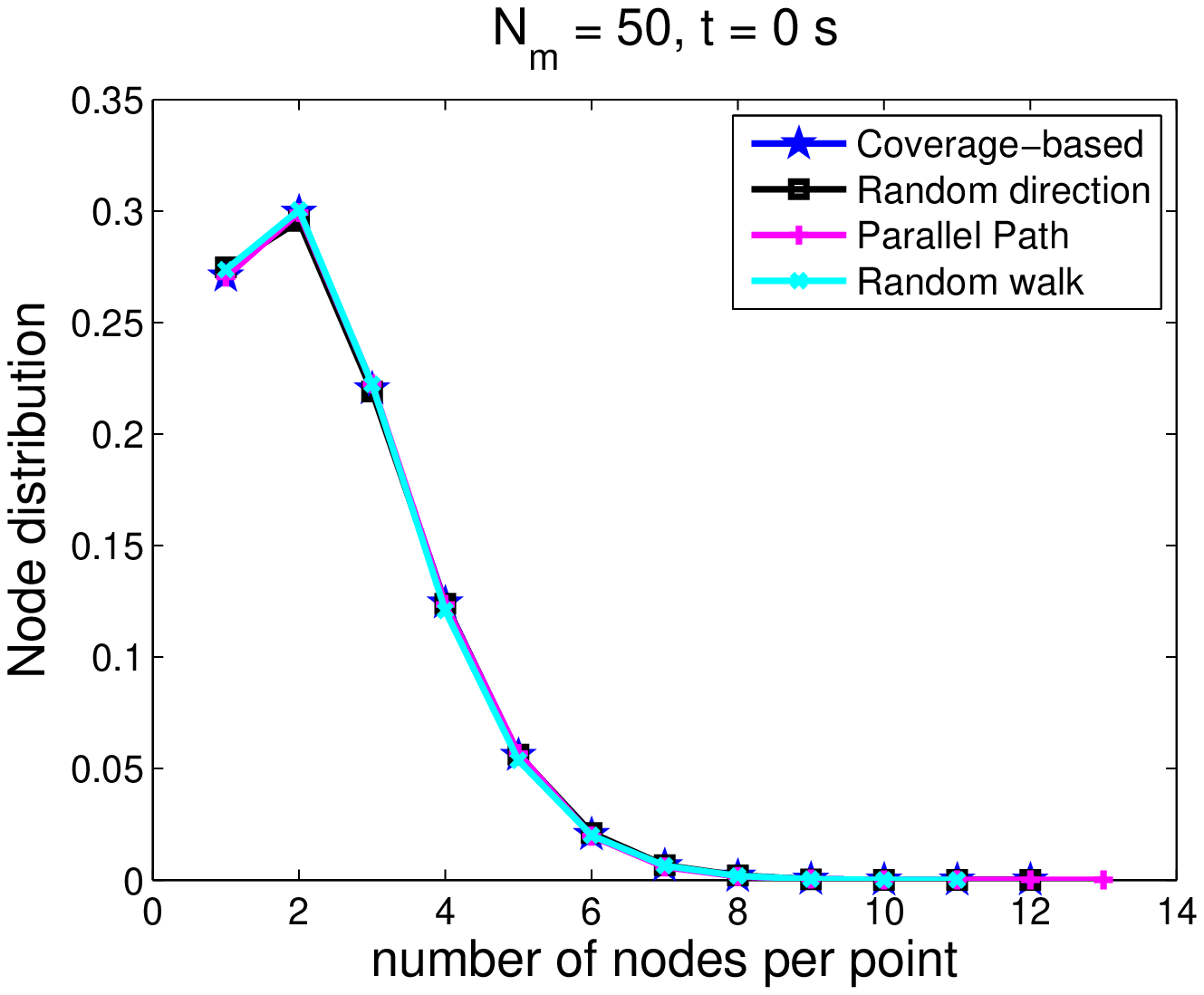,width=0.24\textwidth}}
\subfigure[]
{\psfig{figure=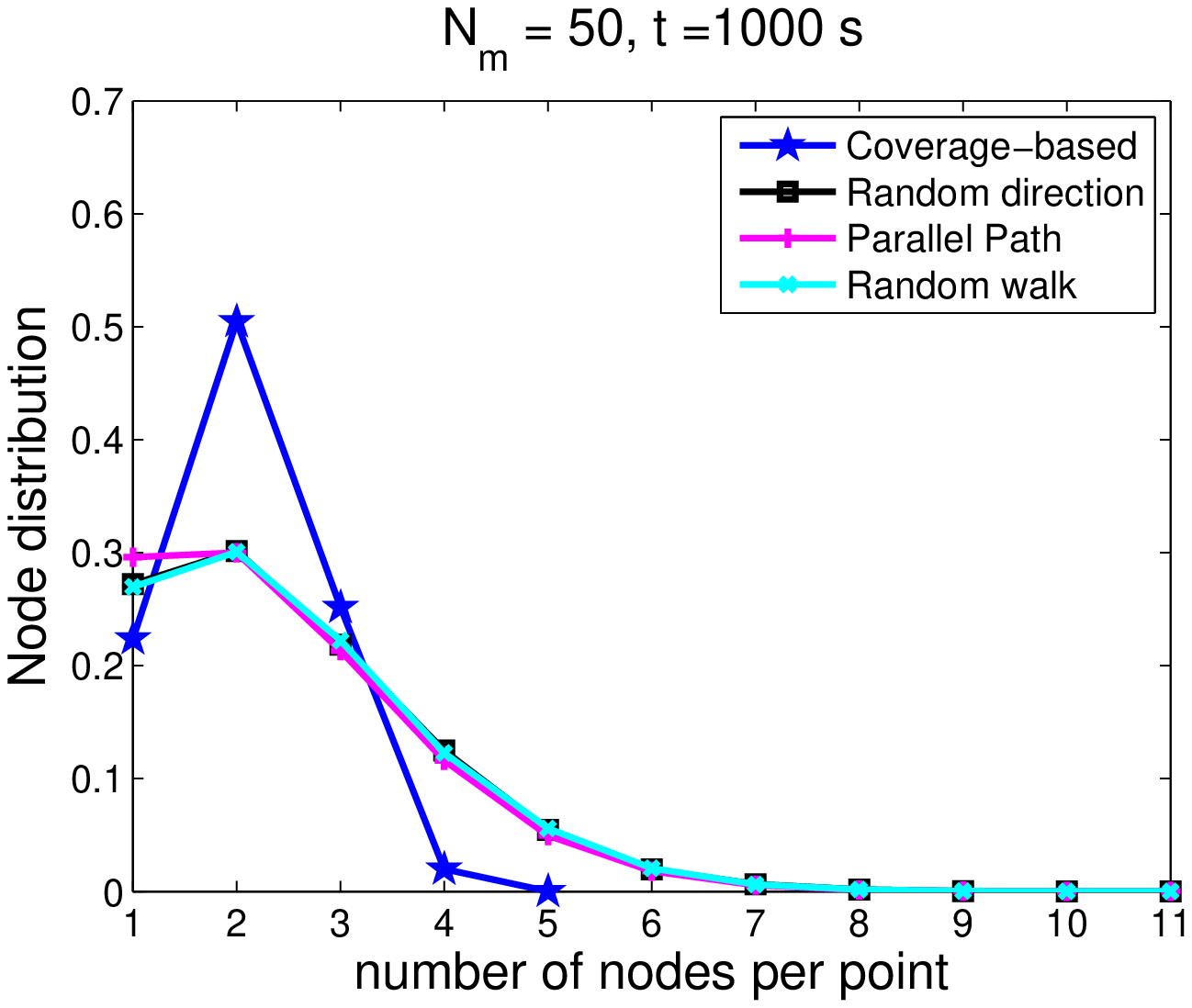,width=0.24\textwidth}}\\
\subfigure[]
{\psfig{figure=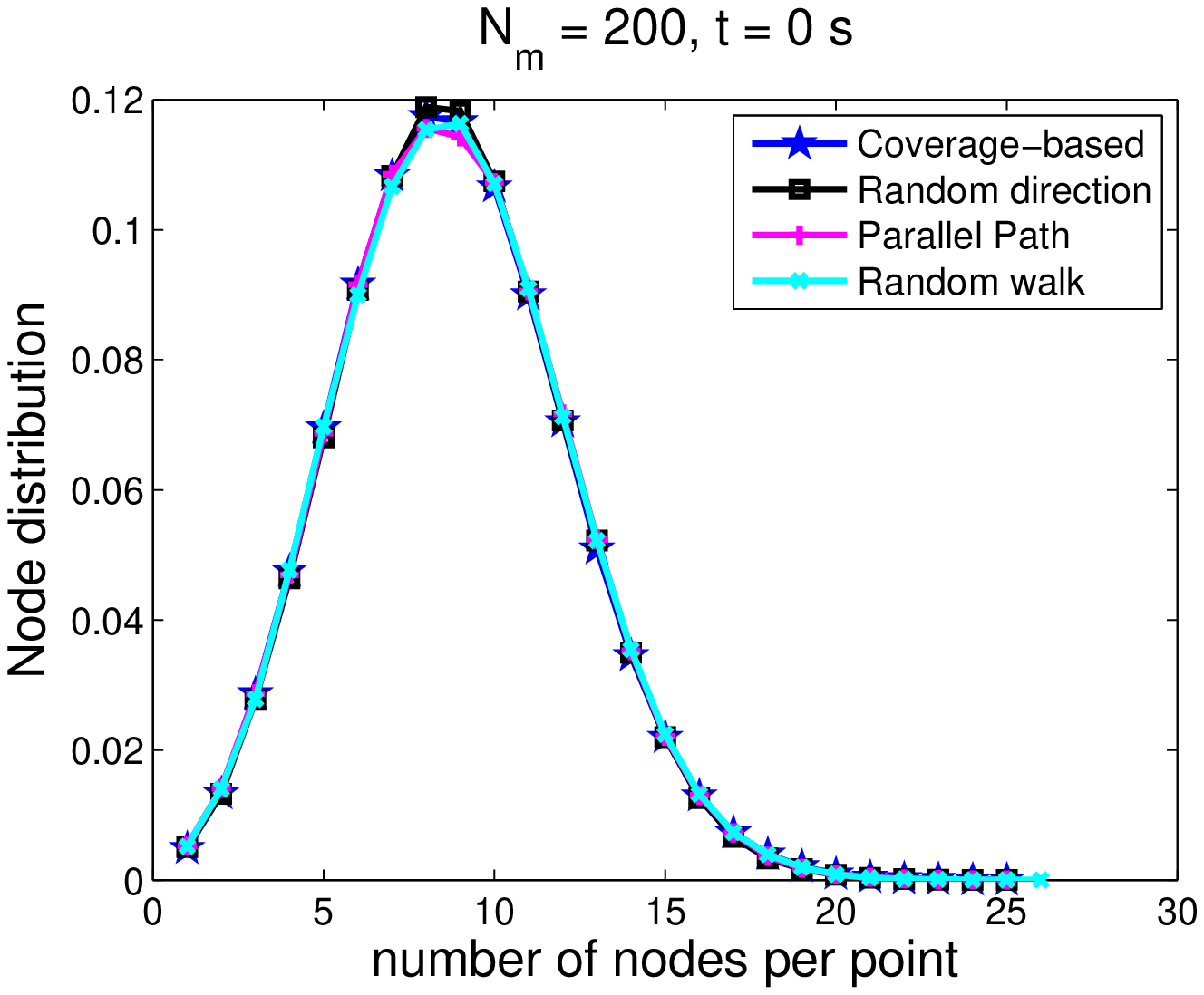,width=0.24\textwidth}}
\subfigure[]
{\psfig{figure=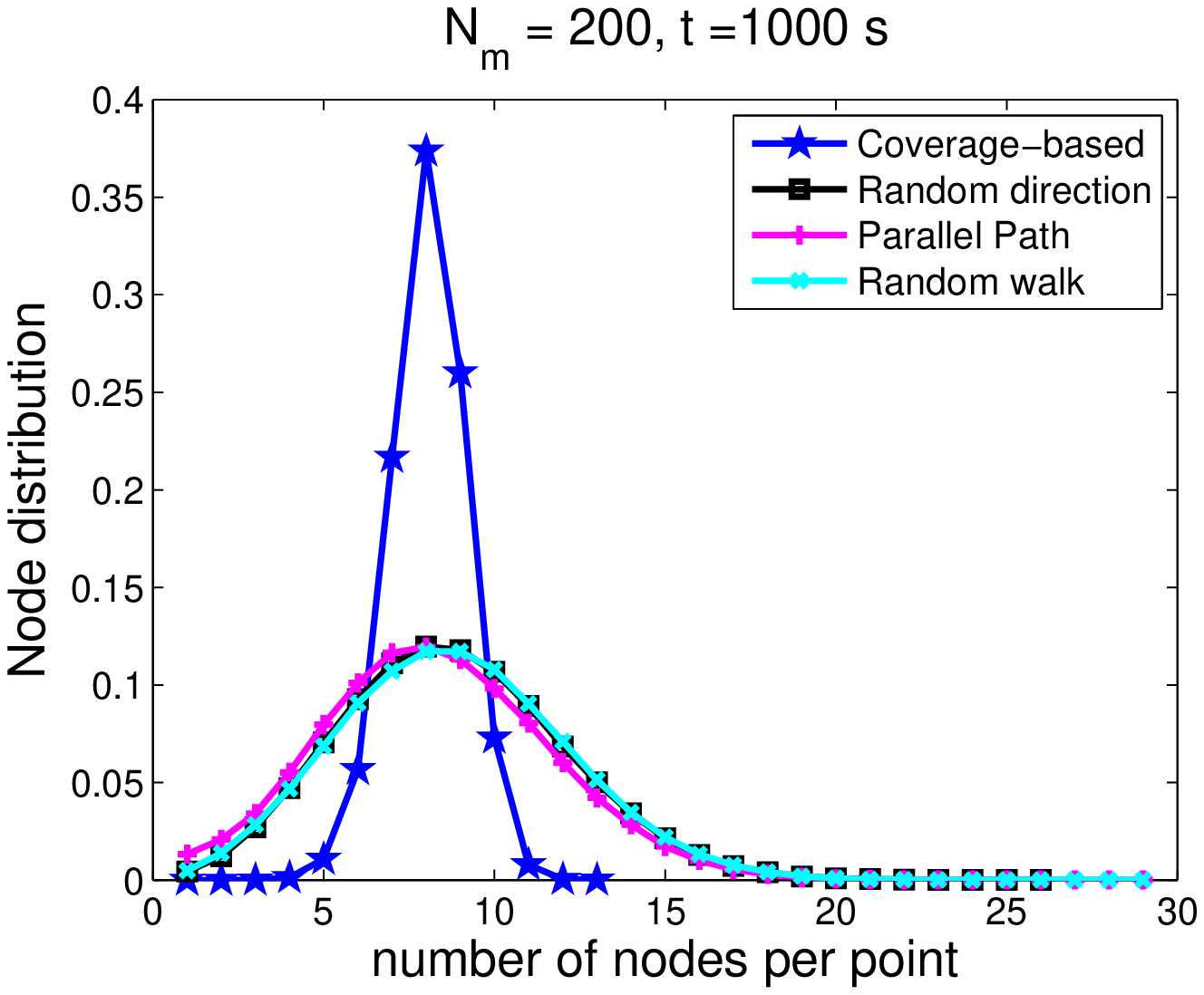,width=0.24\textwidth}}
\caption{Spatial node distribution, when (a) $N_m = 50$ at $t=0$s, (b) $N_m = 50$ at $t=1000$s, (c) $N_m = 200$ at $t=0$s, and (d) $N_m = 200$ at $t=1000$s.}
\label{NodeDist}
\end{figure}

As an illustration, snapshots of spatial node distribution for several mobility models are shown in Fig.~\ref{NodeDist} for $N_m = $ 50 and 200 at times $t = $ 0 and 1000s. Since the initial locations of the mobile nodes are uniformly random, the node distribution of all mobility models are the same at $t=0$s. As time progresses, random walk, random direction, and parallel-path mobility models exhibit the same behavior with an exponential tail. The coverage-based mobility model, on the other hand, results in a shorter tail than the others illustrating the fact that the overlap between the mobile node coverages is much smaller.

\section{Results and Discussion}

In this section, performance comparison of several mobility models in terms of event detection probability and tracking (monitoring) efficiency is provided via Monte Carlo simulations, where each data point is computed over 2000 different runs. It is assumed that the range of the mobile nodes, $r$, is 500m. The simulation area is square-shaped with a length of 4000m. Initially, mobile nodes are randomly distributed in the simulation area. When a mobile node approaches the boundary of the simulation area, a random direction toward the simulation area is assigned for random walk and coverage-based mobility models. The speed of the mobile nodes is assumed to be 5 m/s. The directions of the mobile nodes are updated every 50 m. We assume that a single event occurs at a random location within the simulation area and lasts for a duration of $t_d$ seconds for stationary target case and moves $t_d$ seconds for mobile target case. 

\subsection{Stationary target}

\begin{figure}[!htb]
\centering
\subfigure[]
{\psfig{figure=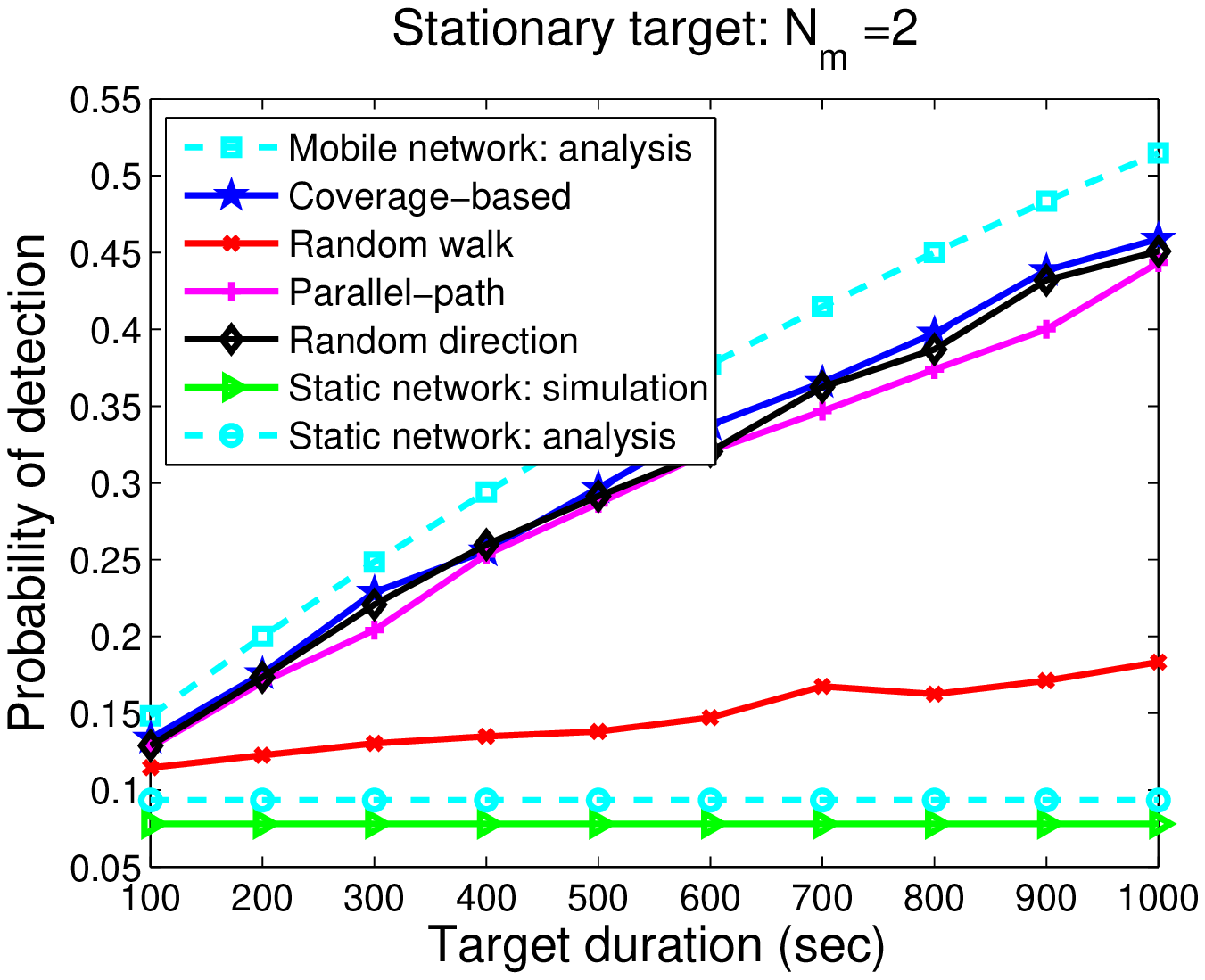,width=0.24\textwidth}}
\subfigure[]
{\psfig{figure=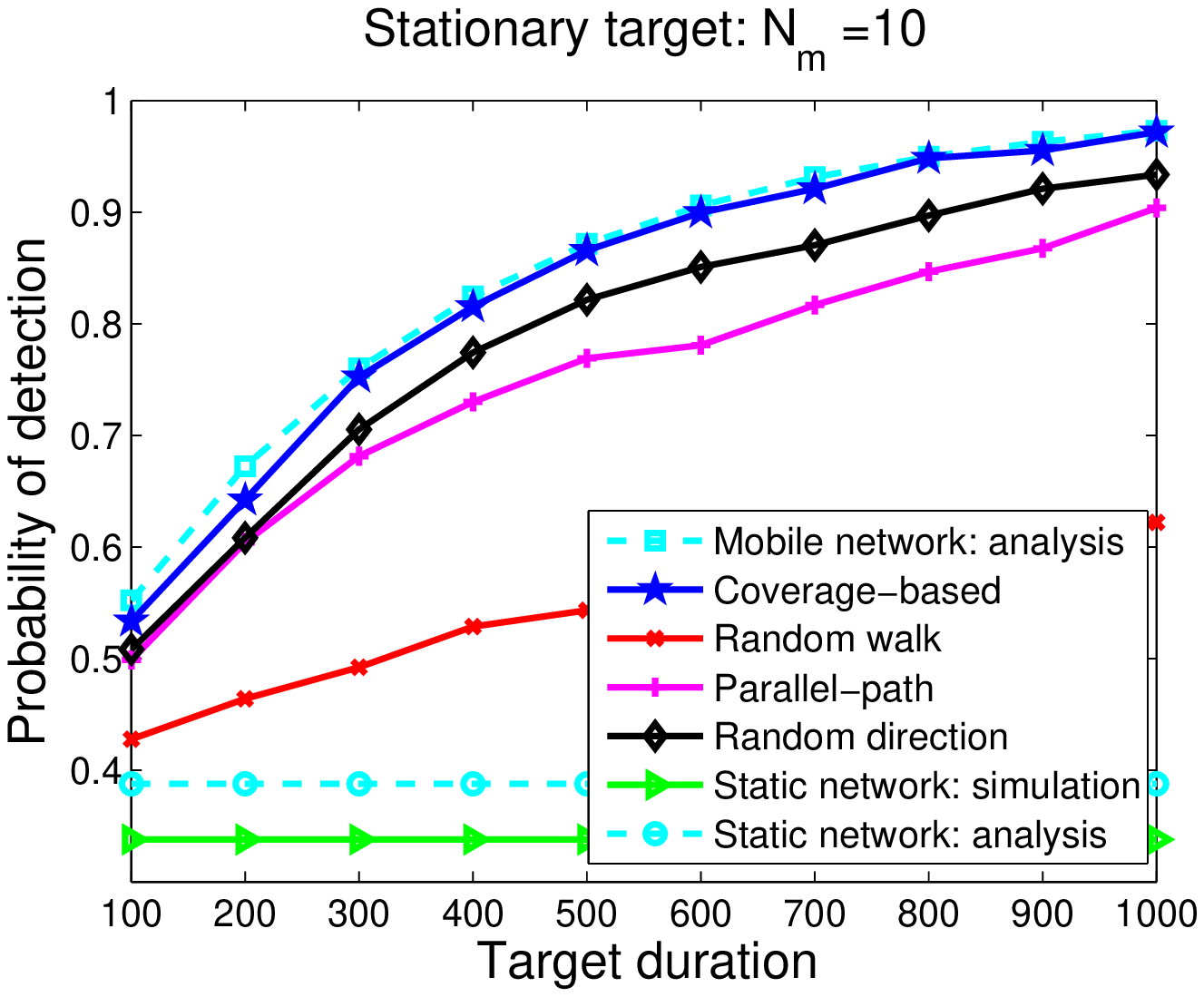,width=0.24\textwidth}}\\
\subfigure[]
{\psfig{figure=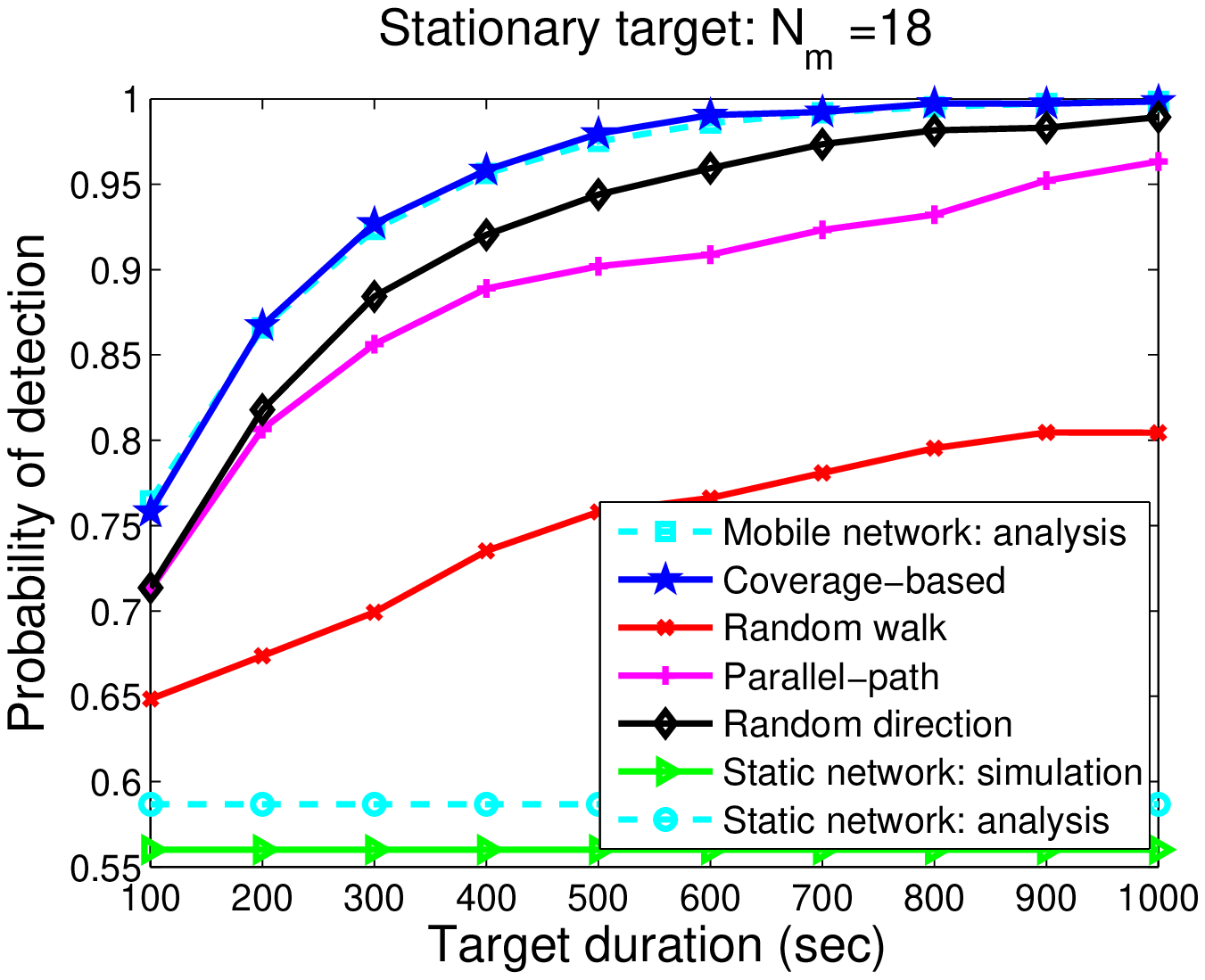,width=0.24\textwidth}}
\subfigure[]
{\psfig{figure=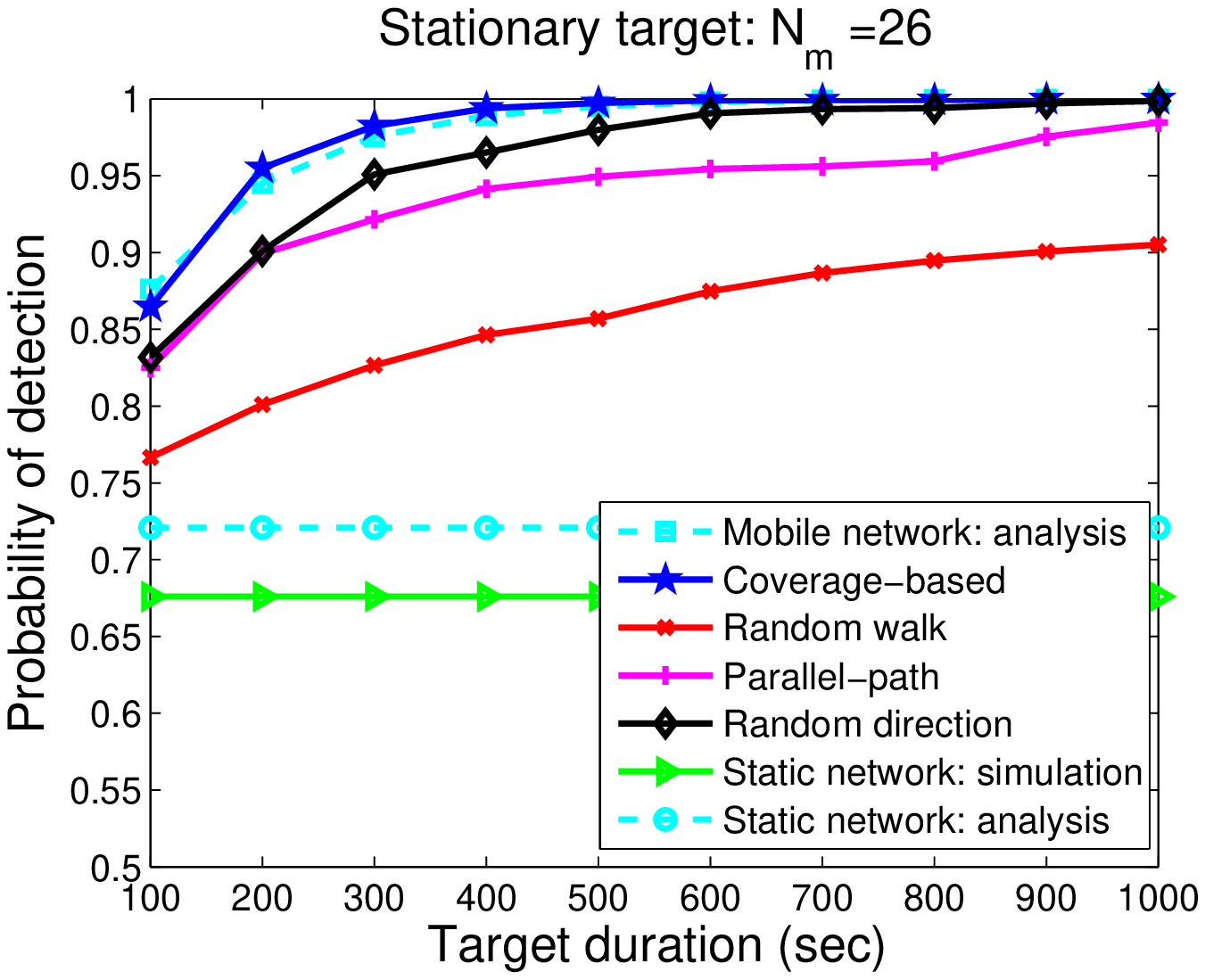,width=0.24\textwidth}}
\caption{{\small Probability of detection versus stationary target duration, when (a) $N_m = 2$, (b) $N_m = 10$, (c) $N_m = 18$, and (d) $N_m = 26$.}}
\label{Pd_vs_td}
\end{figure}

First, we study the impact of target duration and number of mobile nodes on the probability of detection performance of several mobility models. Fig.~\ref{Pd_vs_td} shows the probability of detection versus target duration, when $N_m = \{2,10,18,26\}$. Analytical results are obtained using Eq.'s~(\ref{Pc_anl}) and (\ref{Pcm_anl}). Observe that in all cases random walk model results in the worst performance. While for $N_m = 2$ the rest of the mobility models perform very similarly, as the number of mobile nodes is increased, coverage-based mobility model outperforms the rest of the mobility models under investigation. For comparison, the target detection probability when a uniformly-random distributed static wireless network is used is also shown. While the detection probability of the static network improves with increasing $N_m$, it is significantly lower than that of mobile sensor network. The benefit of mobility can be observed even with short target durations. 

\begin{figure}[!htb]
\centering
\subfigure[]
{\psfig{figure=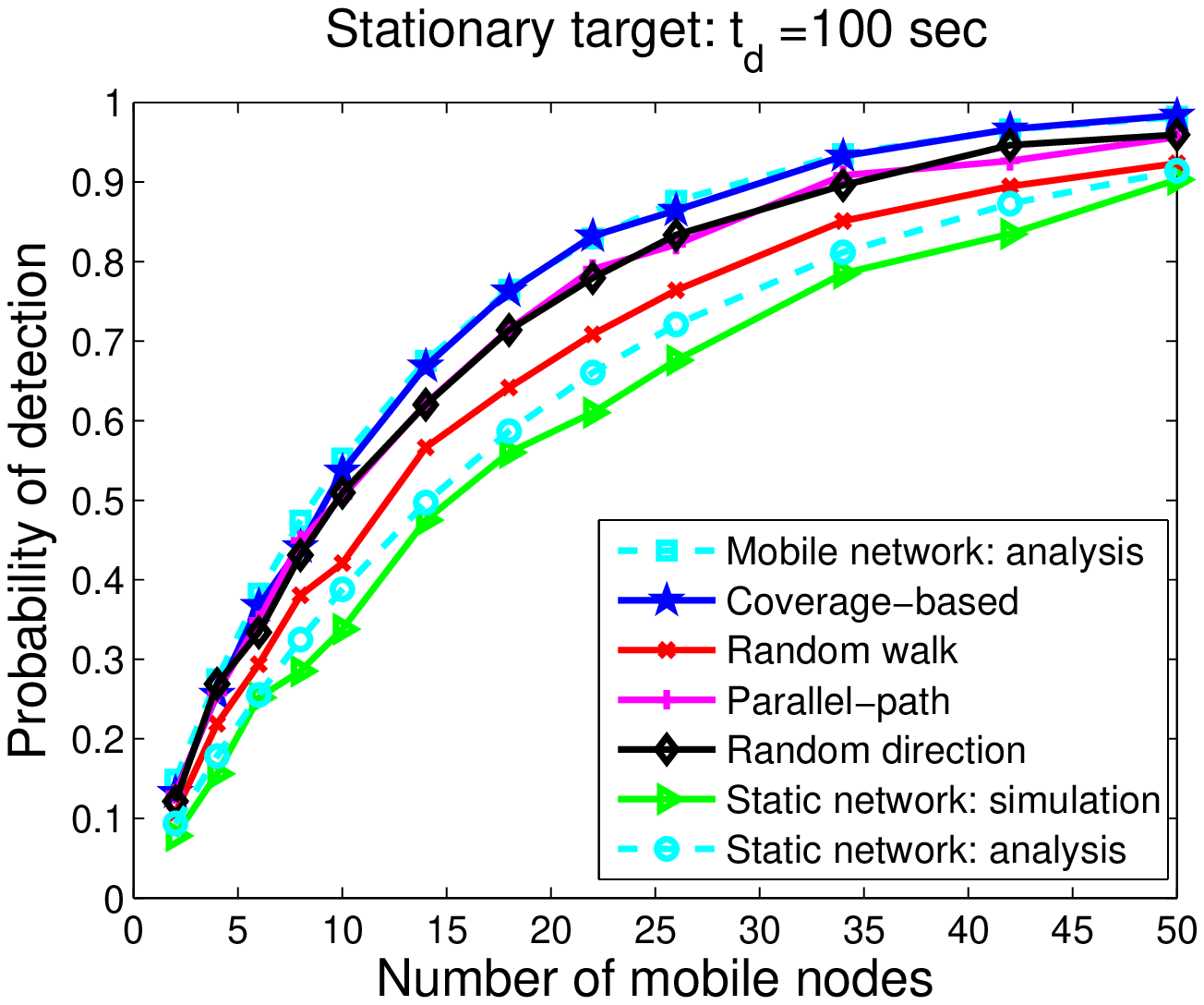,width=0.24\textwidth}}
\subfigure[]
{\psfig{figure=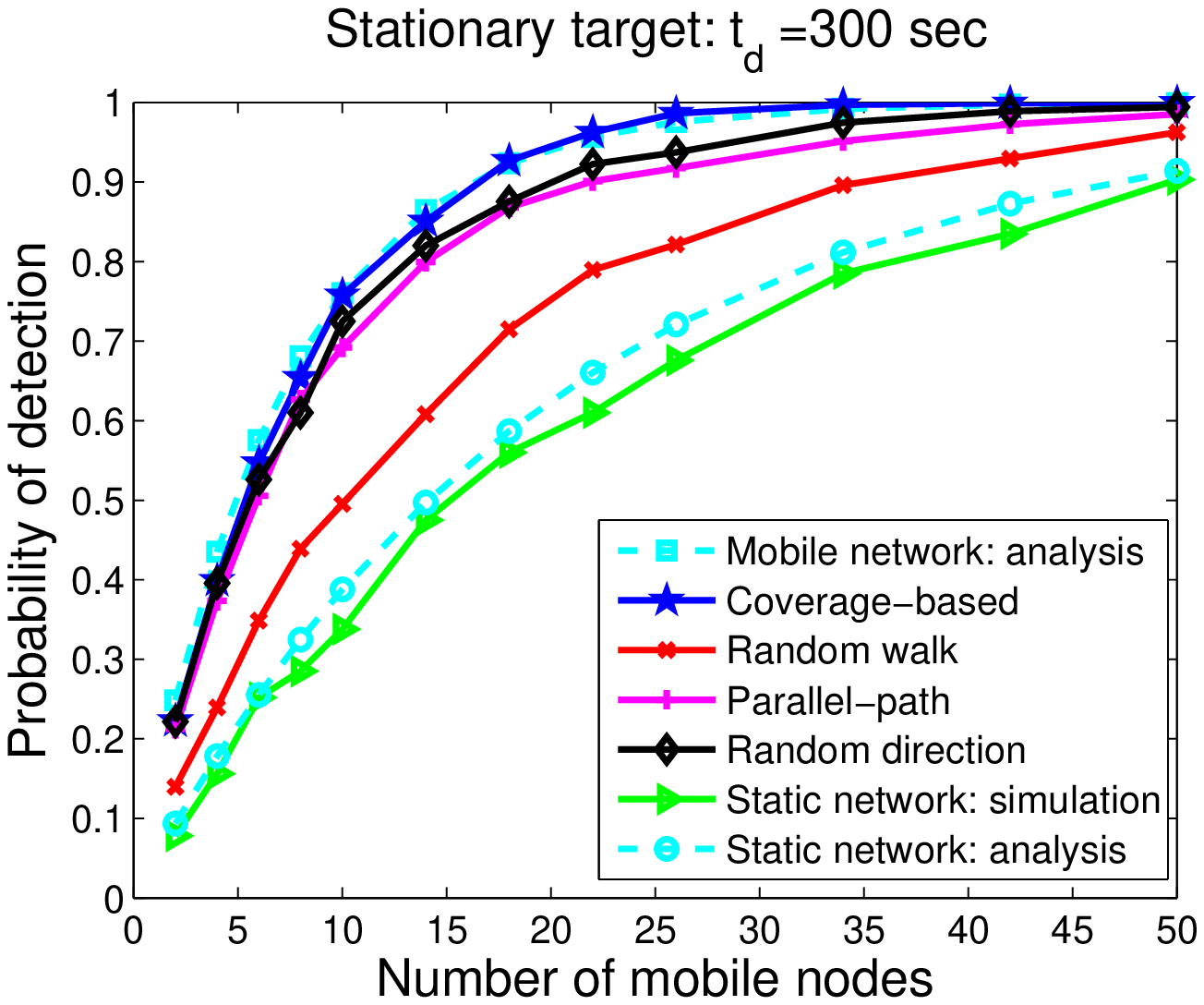,width=0.24\textwidth}}\\
\subfigure[]
{\psfig{figure=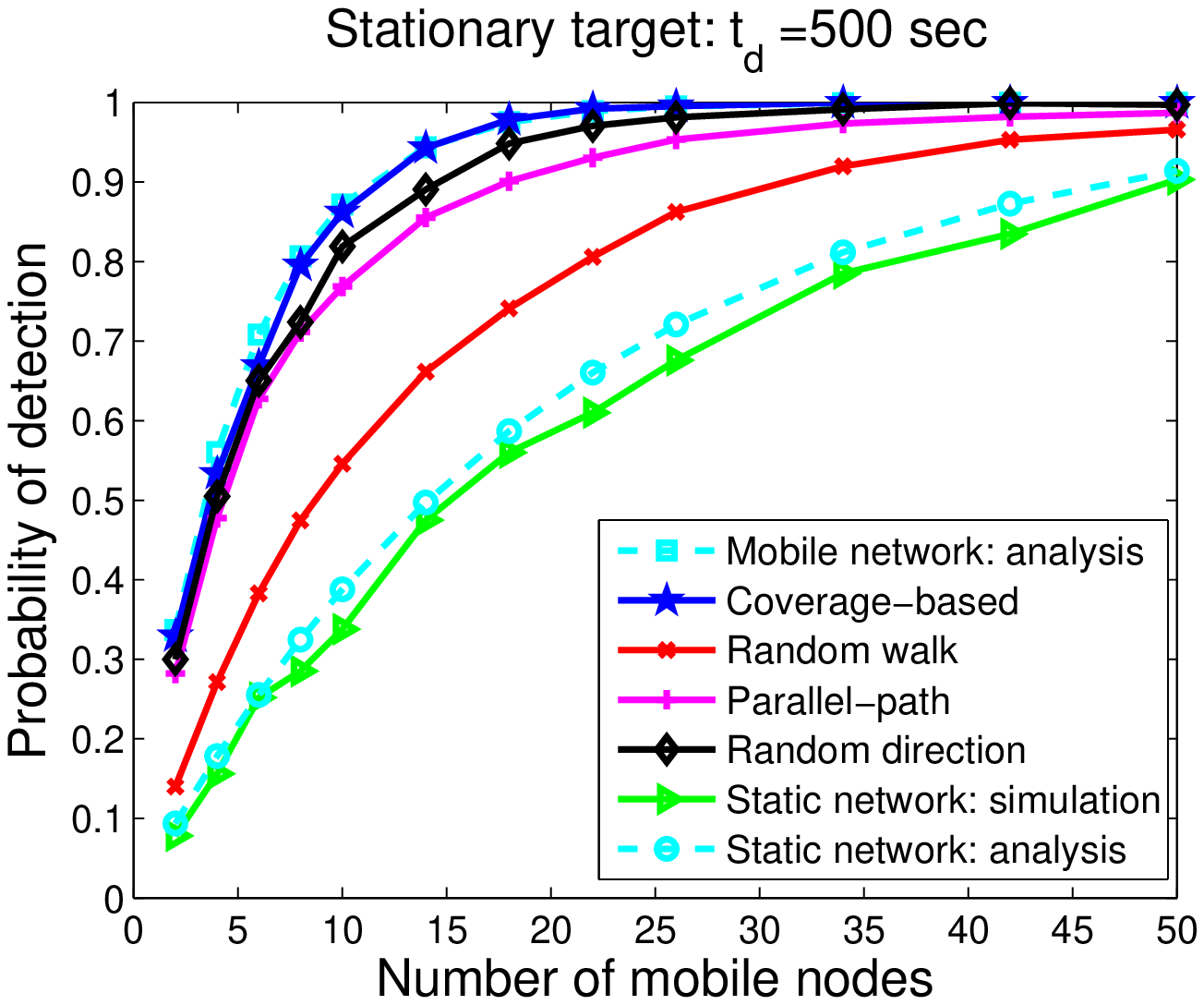,width=0.24\textwidth}}
\subfigure[]
{\psfig{figure=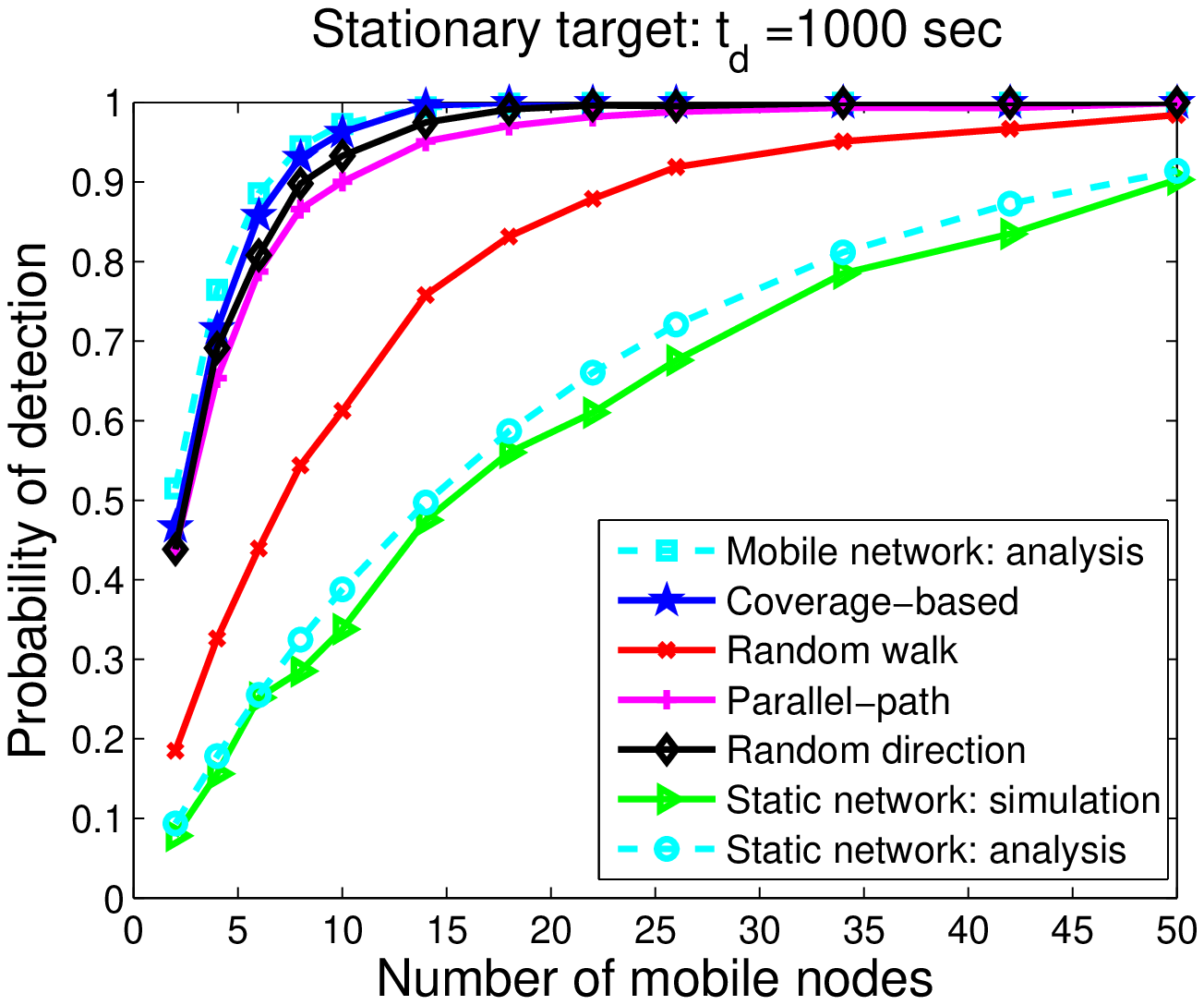,width=0.24\textwidth}}
\caption{Probability of detection versus number of nodes, when (a) $t_d = 100$sec, (b) $t_d = 300$sec, (c) $t_d = 500$sec, and (d) $t_d = 1000$sec .}
\label{Pd_vs_Nm}
\end{figure}

\begin{figure}[!htb]
\centering
\subfigure[]
{\psfig{figure=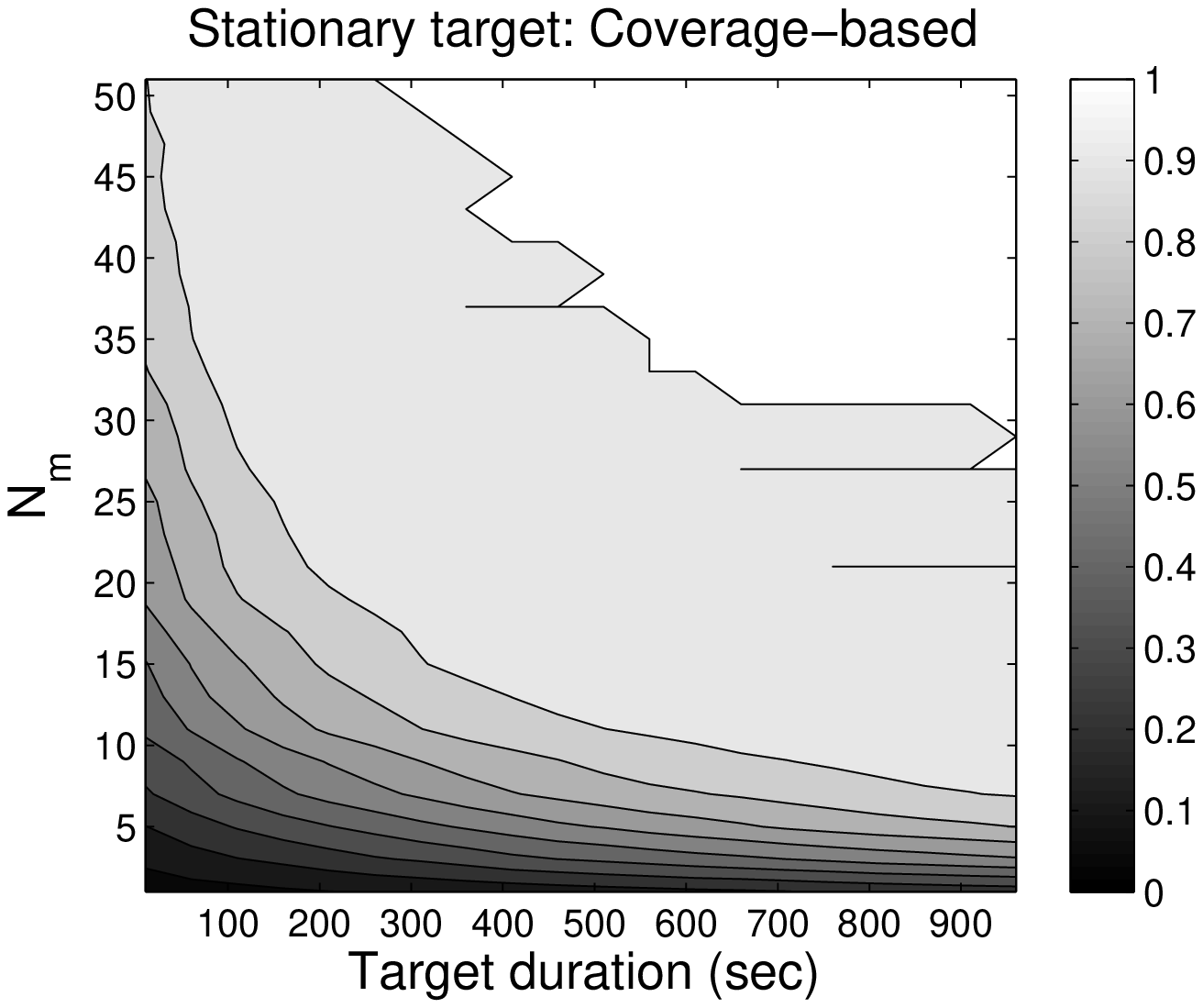,width=0.24\textwidth}}
\subfigure[]
{\psfig{figure=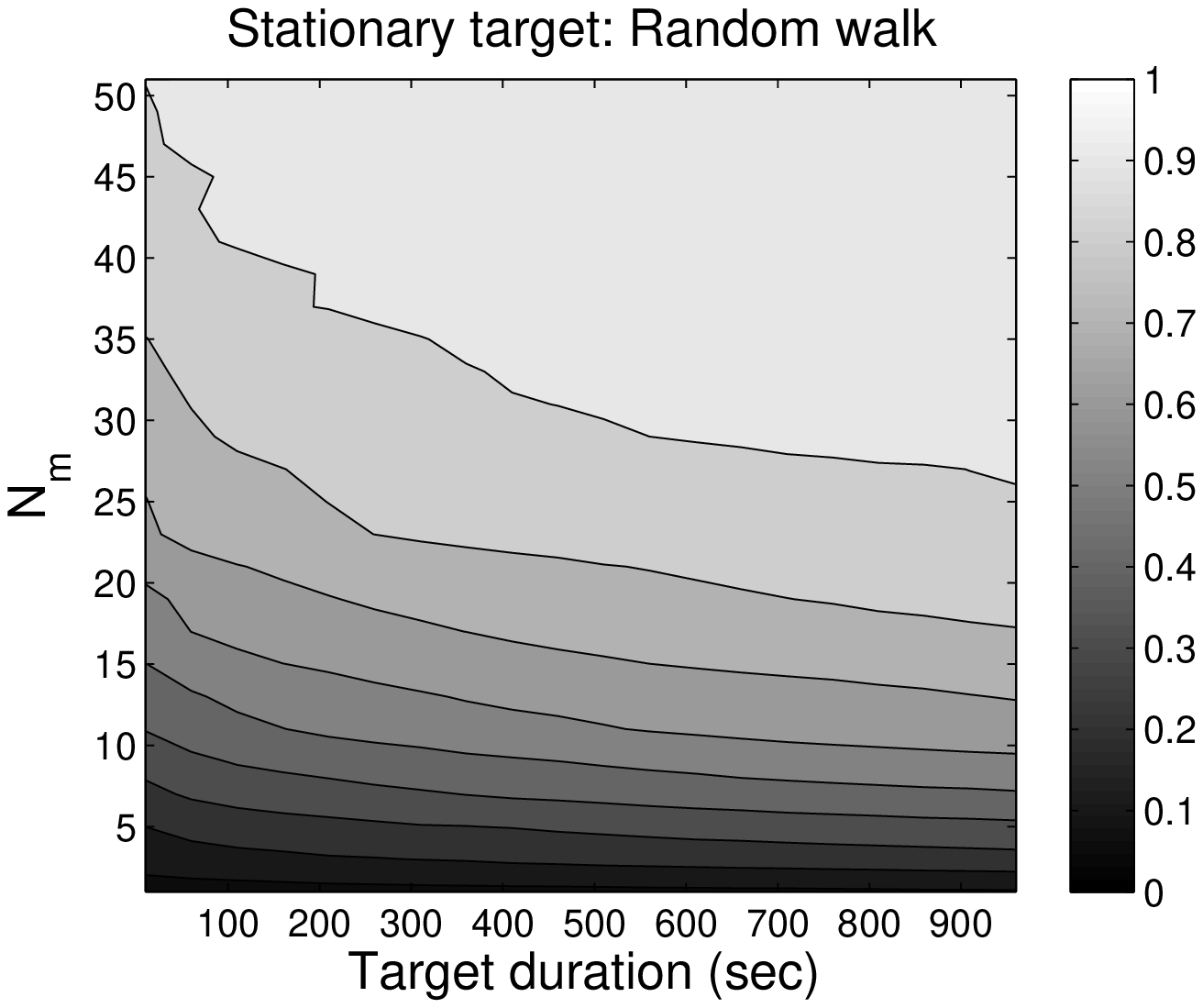,width=0.24\textwidth}}\\
\subfigure[]
{\psfig{figure=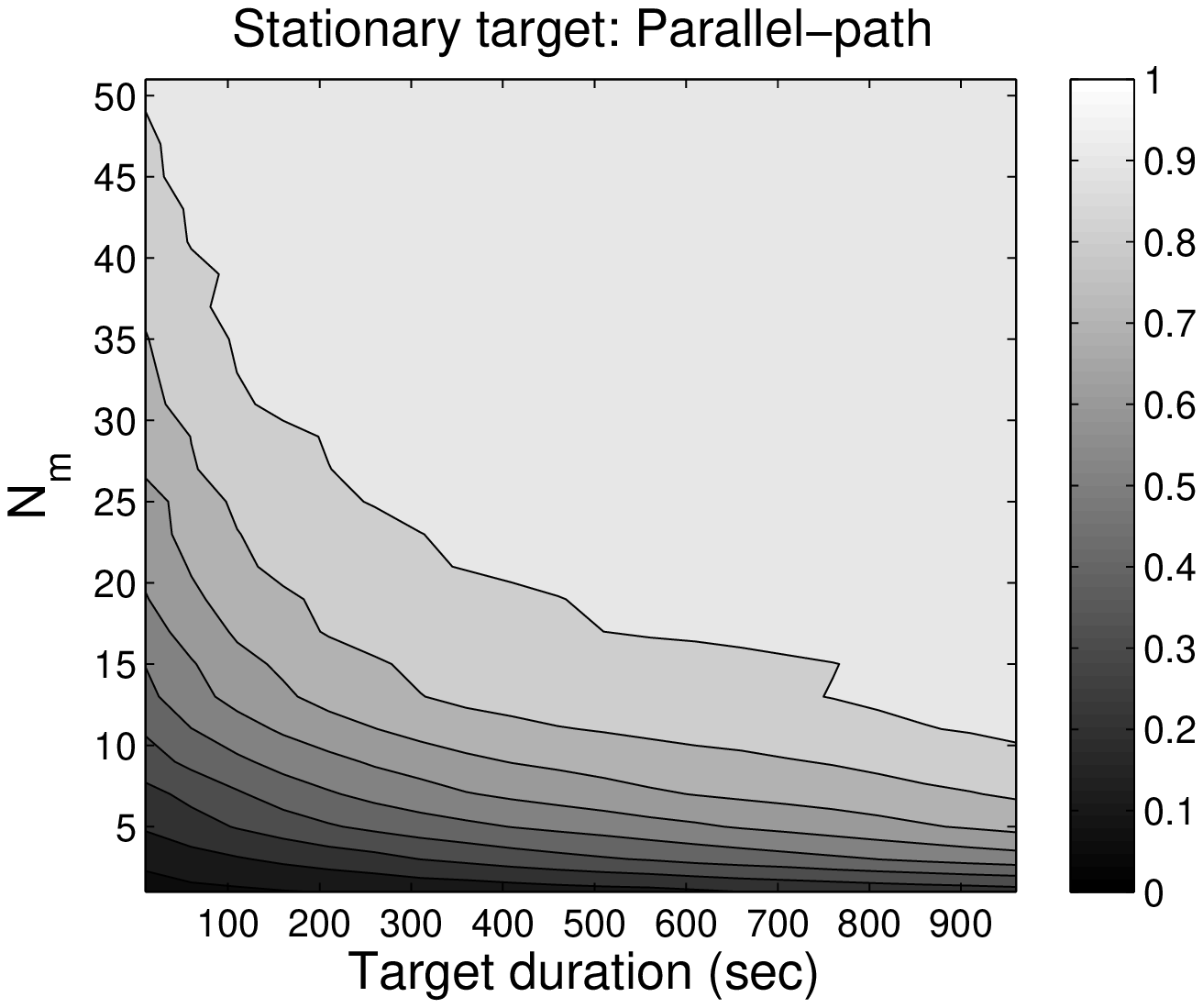,width=0.24\textwidth}}
\subfigure[]
{\psfig{figure=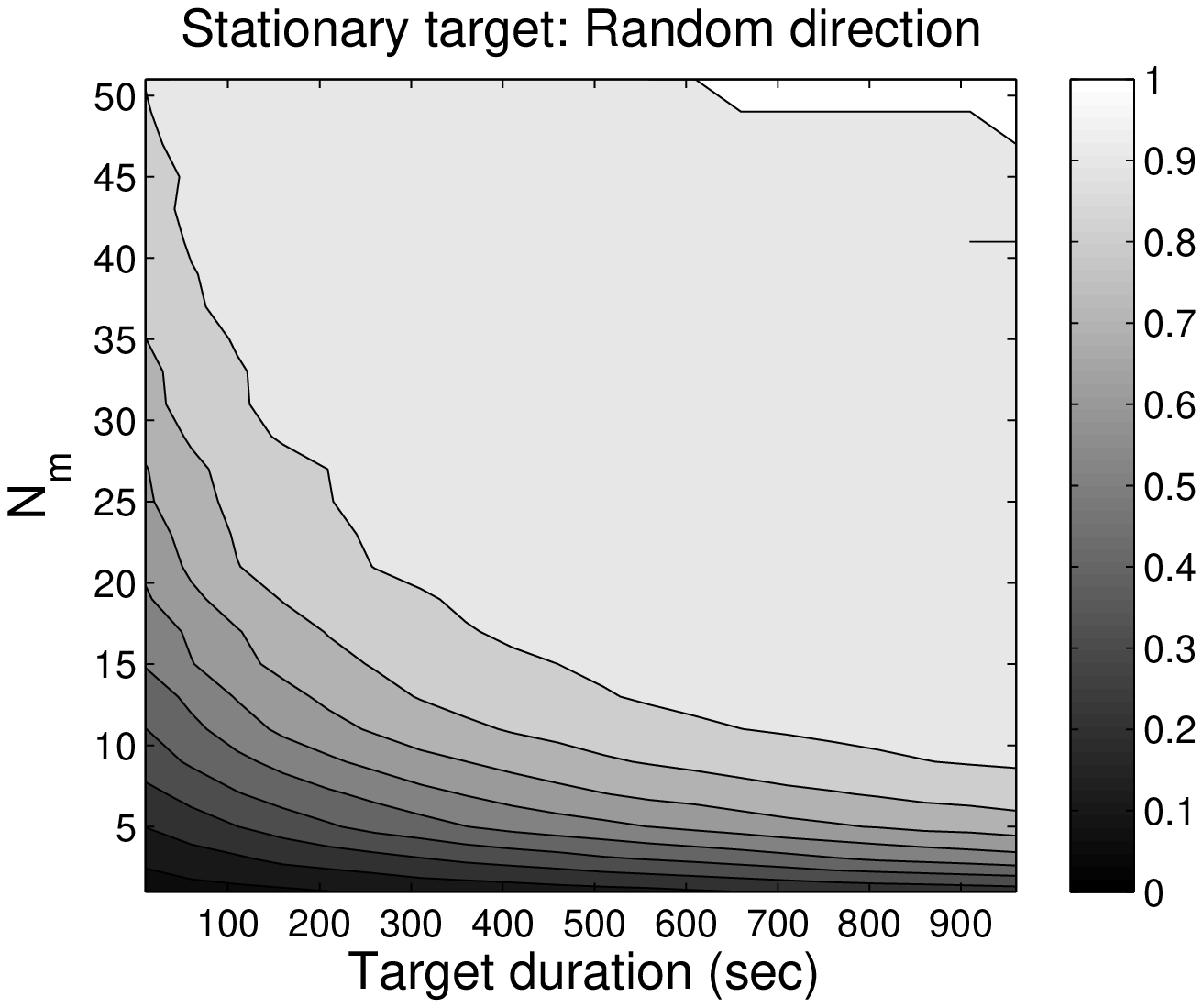,width=0.24\textwidth}}
\caption{Contour plots for probability of detection versus $N_m$ and $t_d$, for (a) Coverage-based, (b) Random walk, (c) Parallel-path, and (d) Random direction mobility models, when the target is stationary.}
\label{Pd_vs_Nm_td}
\end{figure}

Fig.~\ref{Pd_vs_Nm} shows the probability of detection versus $N_m$, when target duration is $\{100,300,500,1000\}$ sec. Similar to the previous case, coverage-based mobility model performs better than the other models. Observe that while the achieved probability of detection values are very close for coverage-based and random direction models, higher probability of detection can be achieved for a wider range of parameter values as shown in Fig.~\ref{Pd_vs_Nm_td}. Observe that the white color, which represents the region where the probability of detection is above 0.95, is significantly larger when coverage-based mobility model is employed. Therefore, if a very high detection probability is required in the deployed wireless sensor network, mobility models other than the coverage-based mobility model might not meet the objective, unless very large number of mobile nodes are deployed.

\begin{figure}[!htb]
\centering
\subfigure[]
{\psfig{figure=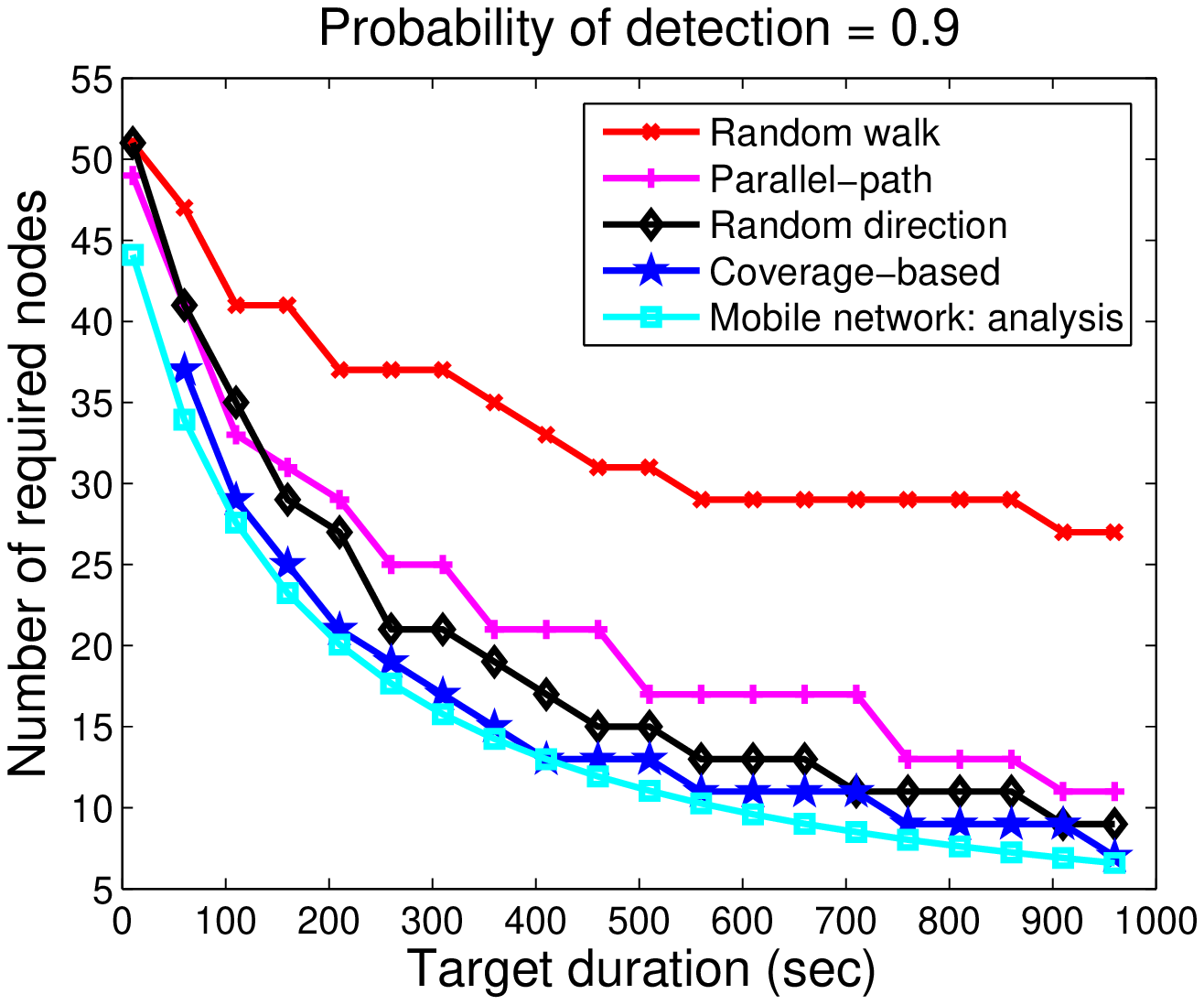,width=0.24\textwidth}}
\subfigure[]
{\psfig{figure=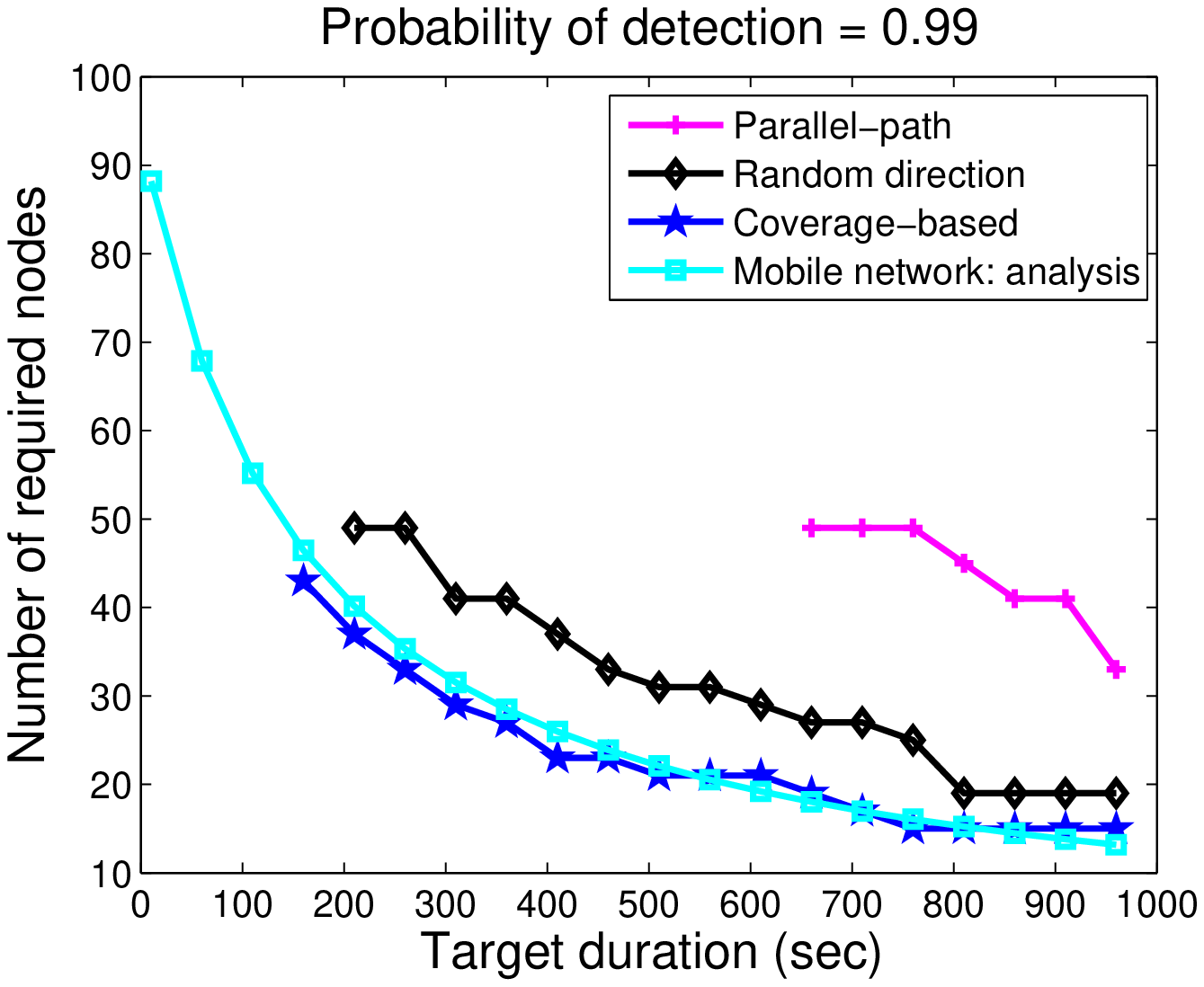,width=0.24\textwidth}}
\caption{Number of required nodes versus target duration, when (a) $P_d = 0.9$, (b) $P_d=0.99$.}
\label{Nmin_vs_td}
\end{figure}

Next, we illustrate the number of required nodes to achieve a certain detection probability. Fig.~\ref{Nmin_vs_td} shows the analytical and simulation results for the number of required nodes versus target duration, when the desired probability of detection value is set to $\{0.9,0.99\}$. Observe that coverage-based model performs the best and random walk model performs the worst once more. Coverage-based model shows a great match to the analytical results obtained from Eq.~(\ref{Nmmin_anl}), whereas the other models significantly deviate from the bound as the detection requirements become more stringent. For example, approximately 10 less mobile nodes are required to achieve a detection probability of 0.99 with coverage-based mobility model than the nearest, i.e., random direction model. 

\subsection{Mobile Target: Linear motion}

In this section, we investigate the target detection and tracking performance of static and mobile wireless sensor networks. In addition to detection probability we also compute the percentage tracking time for several scenarios, where percentage tracking time is defined as the ratio of the time the target is within the coverage of at least one sensor node to total duration of event. First, we assume that the target starts its motion from a randomly selected point in one boundary of the simulation area and moves toward a randomly selected point in the opposite boundary following a line. This scenario could be considered an example of border monitoring, where the target tries to cross the border without being detected.
\begin{figure}[!htb]
\centering
\subfigure[]
{\psfig{figure=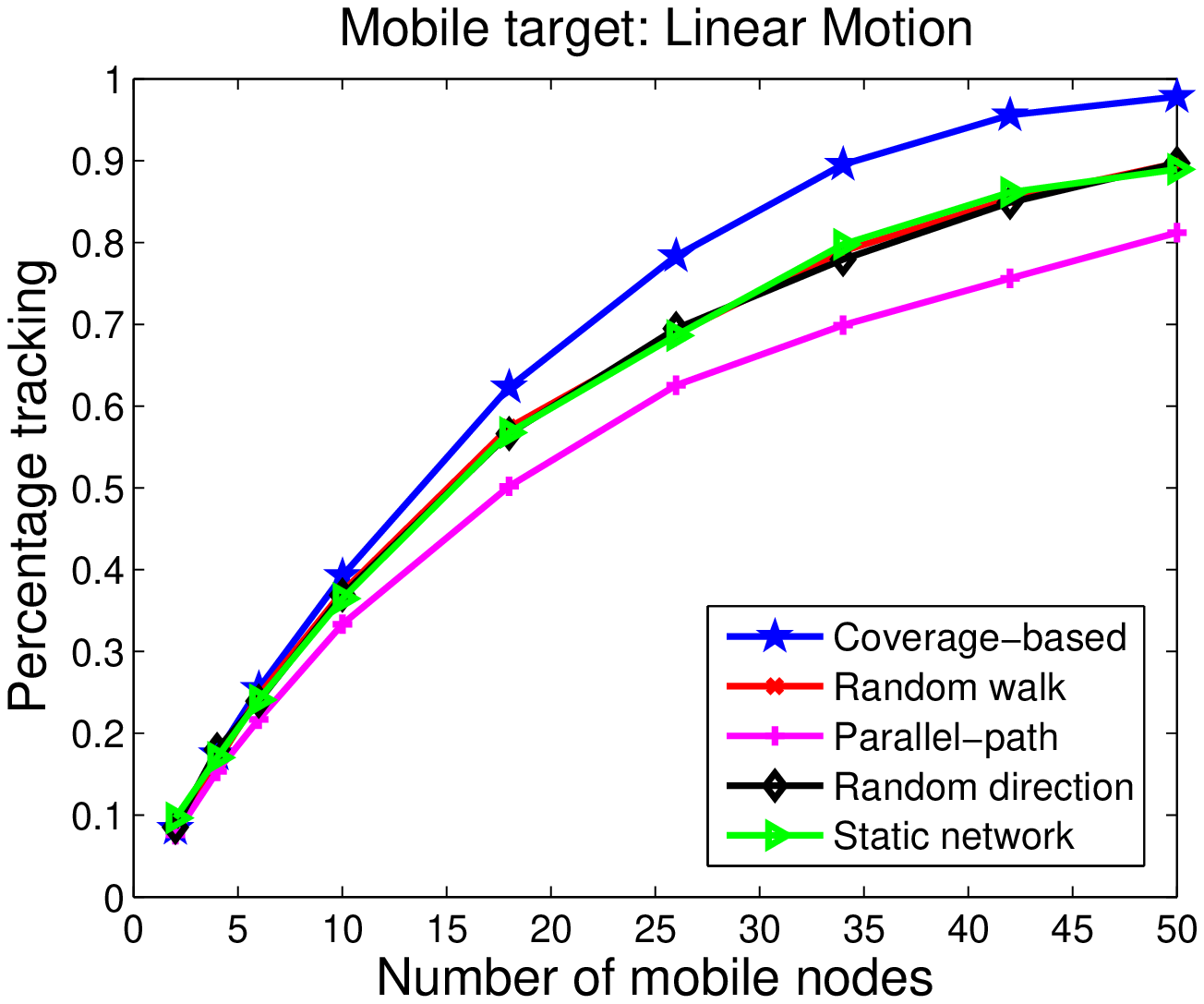,width=0.24\textwidth}}
\subfigure[]
{\psfig{figure=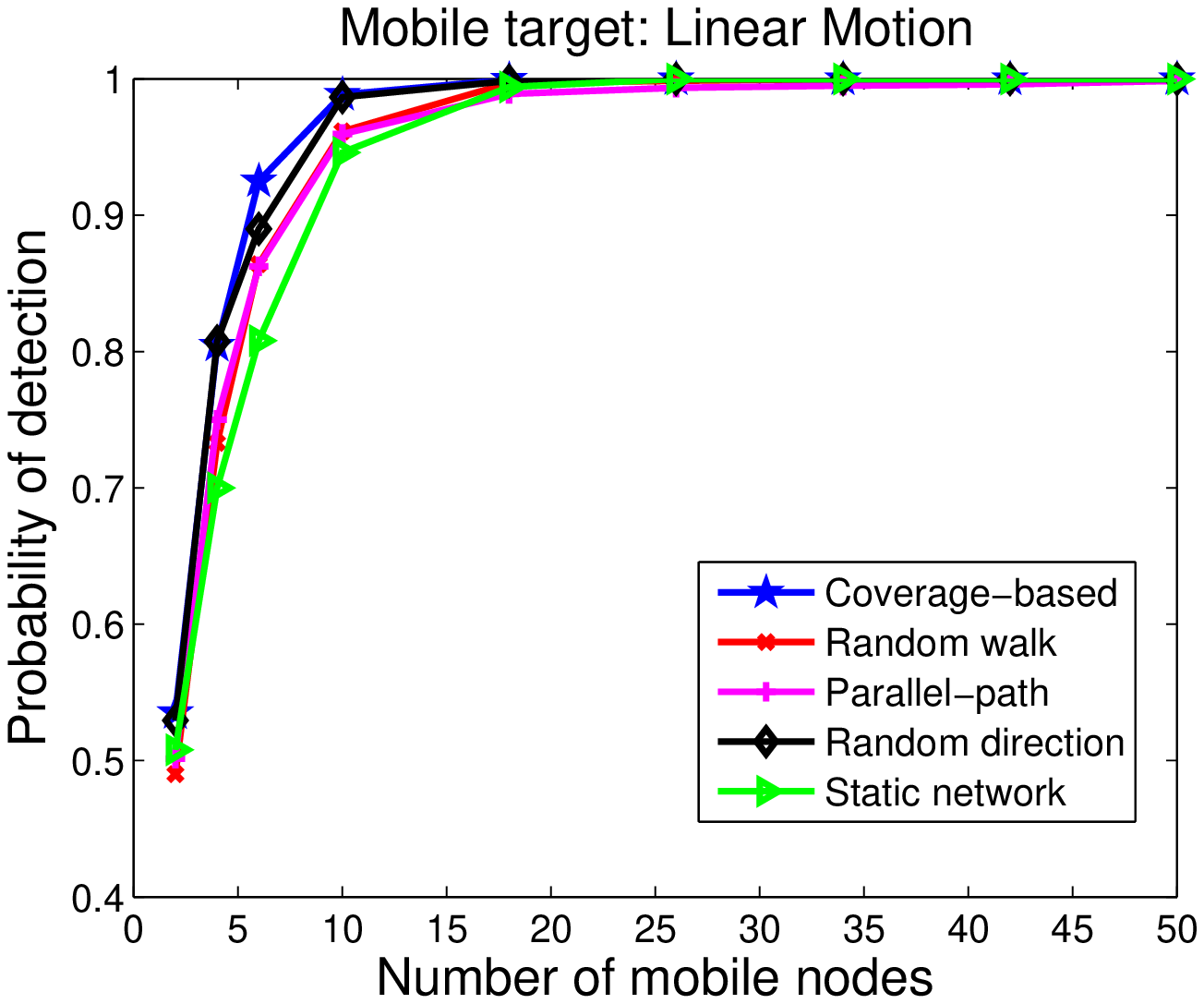,width=0.24\textwidth}}
\caption{(a) Percentage tracking and (b) probability of detection versus $N_m$, for a mobile target moving border-to-border on a linear path.}
\label{Pt_Pdvs_Nm_Lin}
\end{figure}

Fig.~\ref{Pt_Pdvs_Nm_Lin} (a) and (b) show percentage tracking and detection probability performance of the mobile and static wireless sensor networks versus number of mobile nodes. While the detection probability performance of all models are very close to each other, percentage tracking performance of coverage-based mobility model is higher than the others. This is encouraging since the target can be ``tracked'' with few number of sensor nodes even though whole geographical area is not covered 100\% of the time. The tracking performance can clearly be improved if the target location information is incorporated into the mobility path after it is detected. However, since the application (i.e., objective) of the sensor network is not specified in this paper, such mobility path design is beyond the scope of this work.

\subsection{Mobile Target: Random walk}

\begin{figure}[!htb]
\centering
\subfigure[]
{\psfig{figure=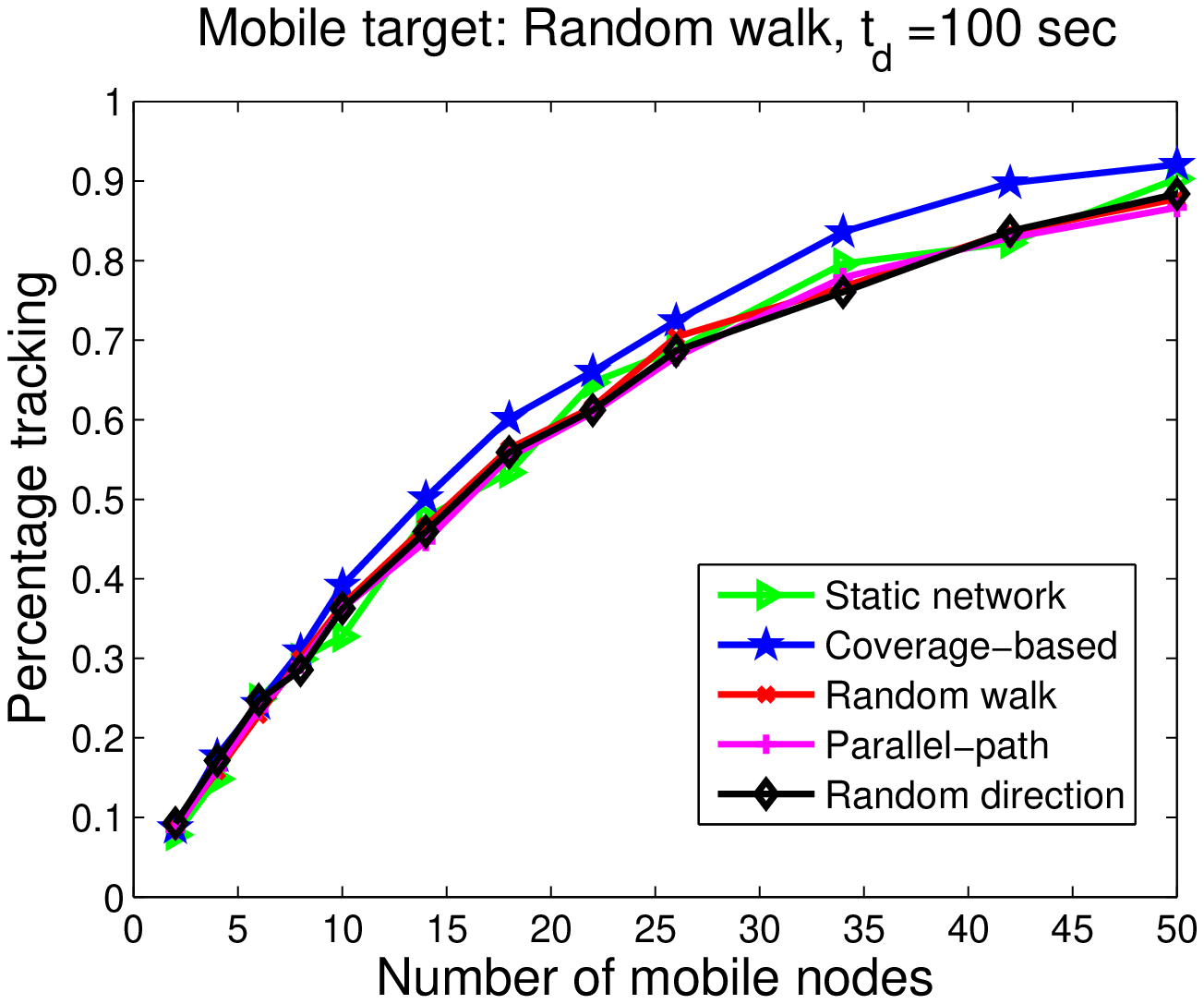,width=0.24\textwidth}}
\subfigure[]
{\psfig{figure=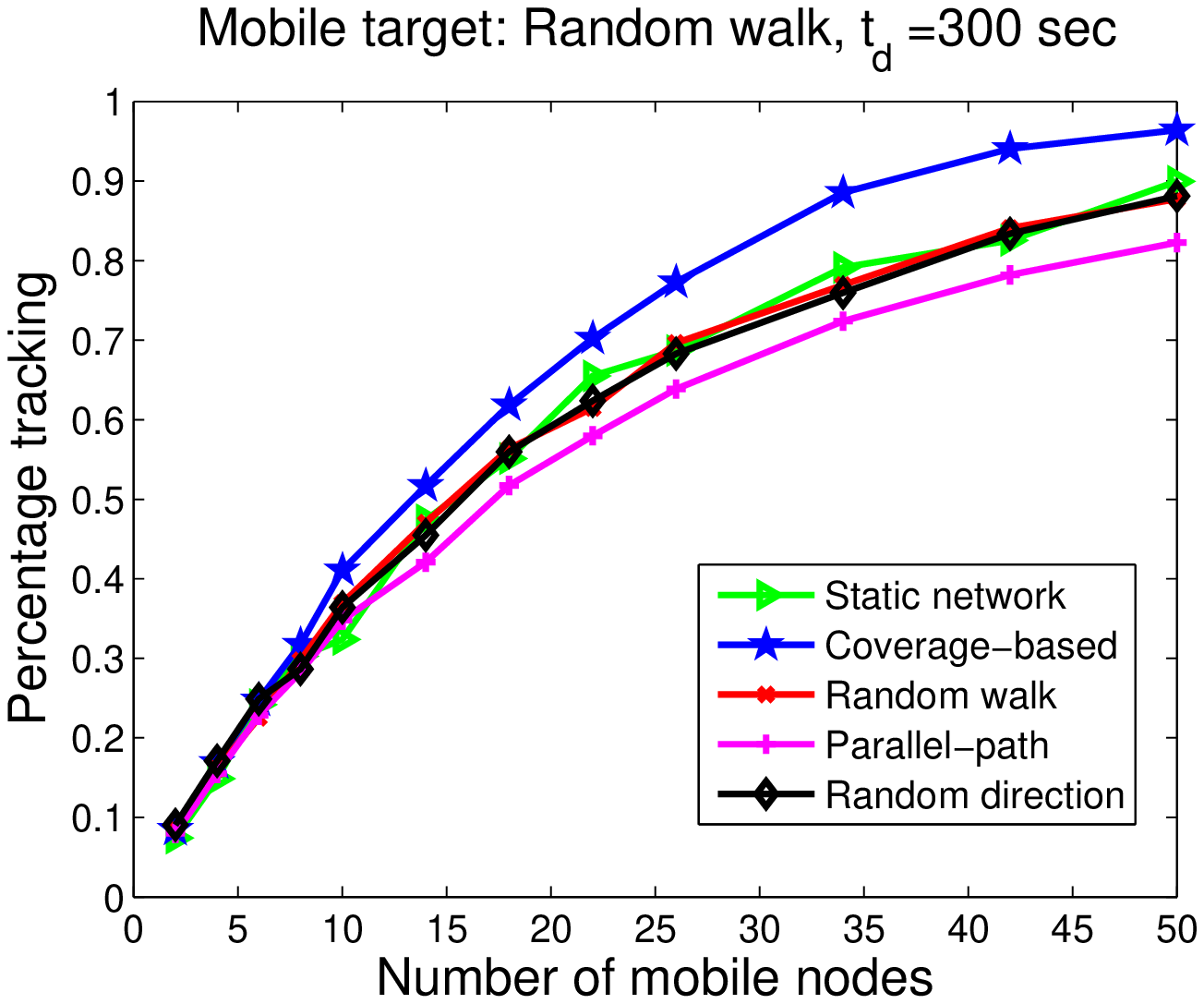,width=0.24\textwidth}}\\
\subfigure[]
{\psfig{figure=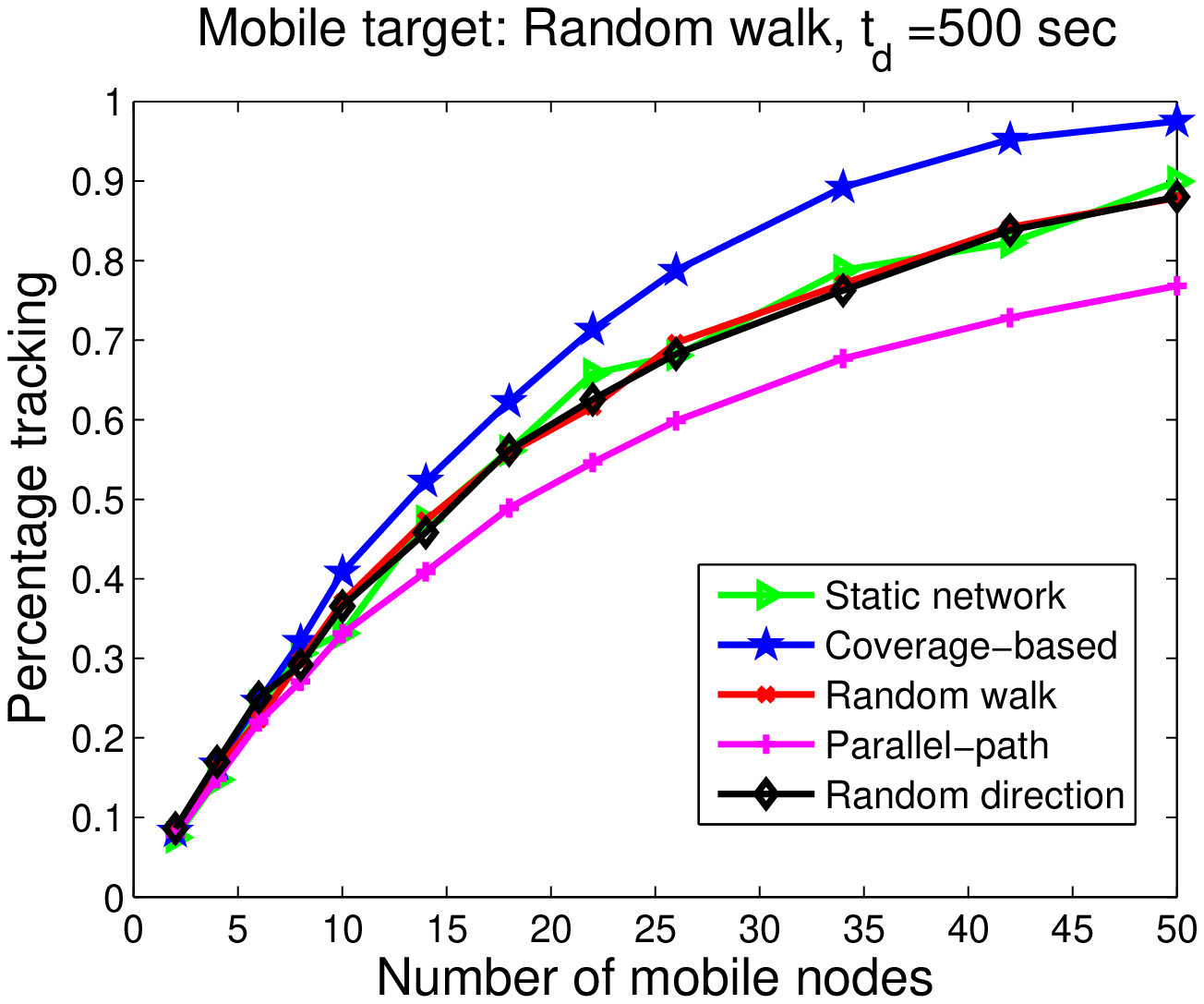,width=0.24\textwidth}}
\subfigure[]
{\psfig{figure=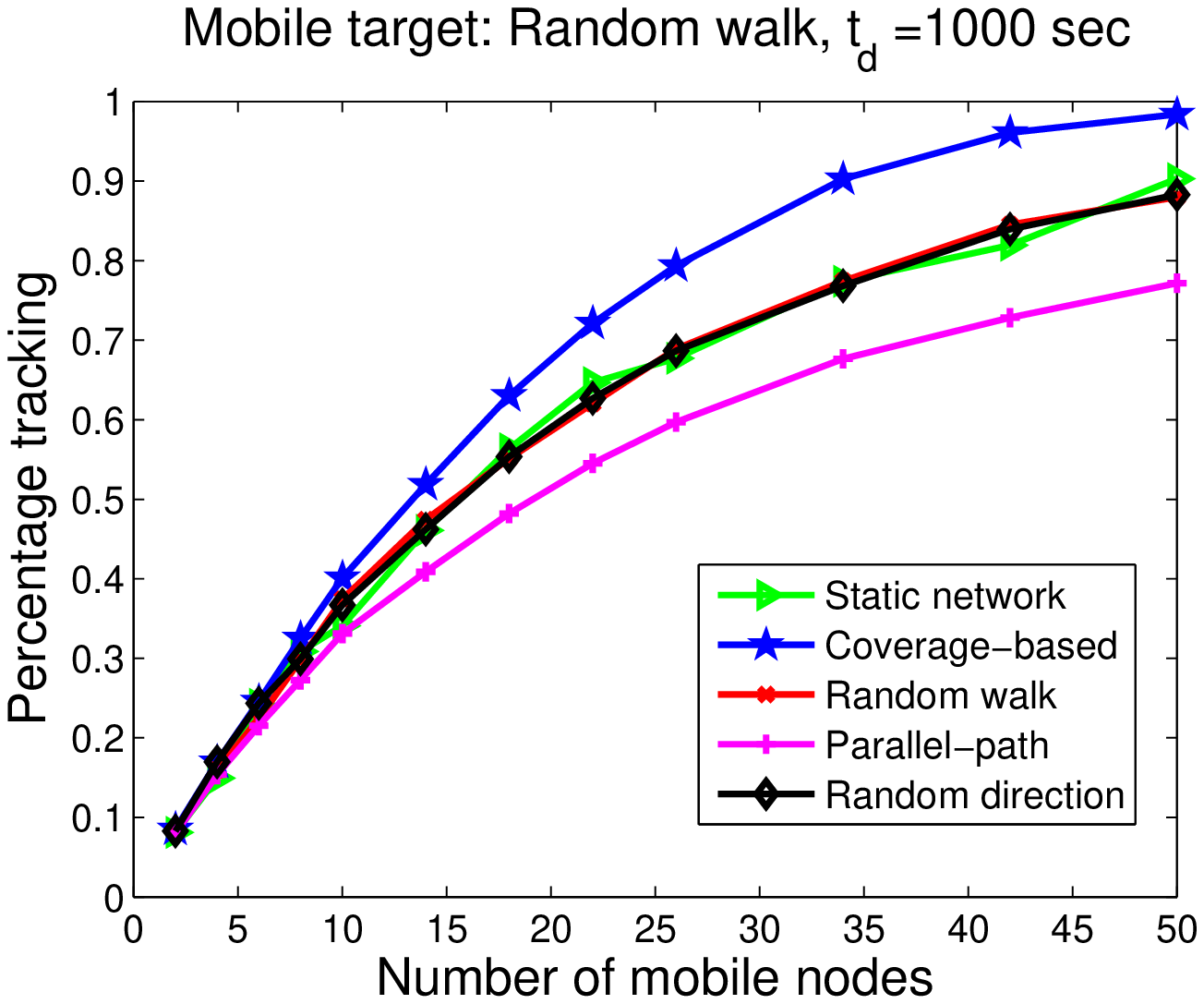,width=0.24\textwidth}}
\caption{Percentage tracking versus number of nodes, when (a) $t_d = 100$sec, (b) $t_d = 300$sec, (c) $t_d = 500$sec, (d) $t_d = 1000$sec.}
\label{Pt_vs_Nm_RW}
\end{figure}

Next, we assume that the target follows a random walk in the geographical area (such as an animal wandering around in its habitat). Fig.~\ref{Pt_vs_Nm_RW} shows the tracking percentage versus number of mobile nodes when $t_d = \{100,300,500,1000\}$ sec. Observe that tracking percentage does not depend on the target duration, whereas the percentage increases with increasing $N_m$ as expected. The tracking percentage performance in this case is very similar to the previous one where the target follows a linear path. 

\begin{figure}[!htb]
\centering
\subfigure[]
{\psfig{figure=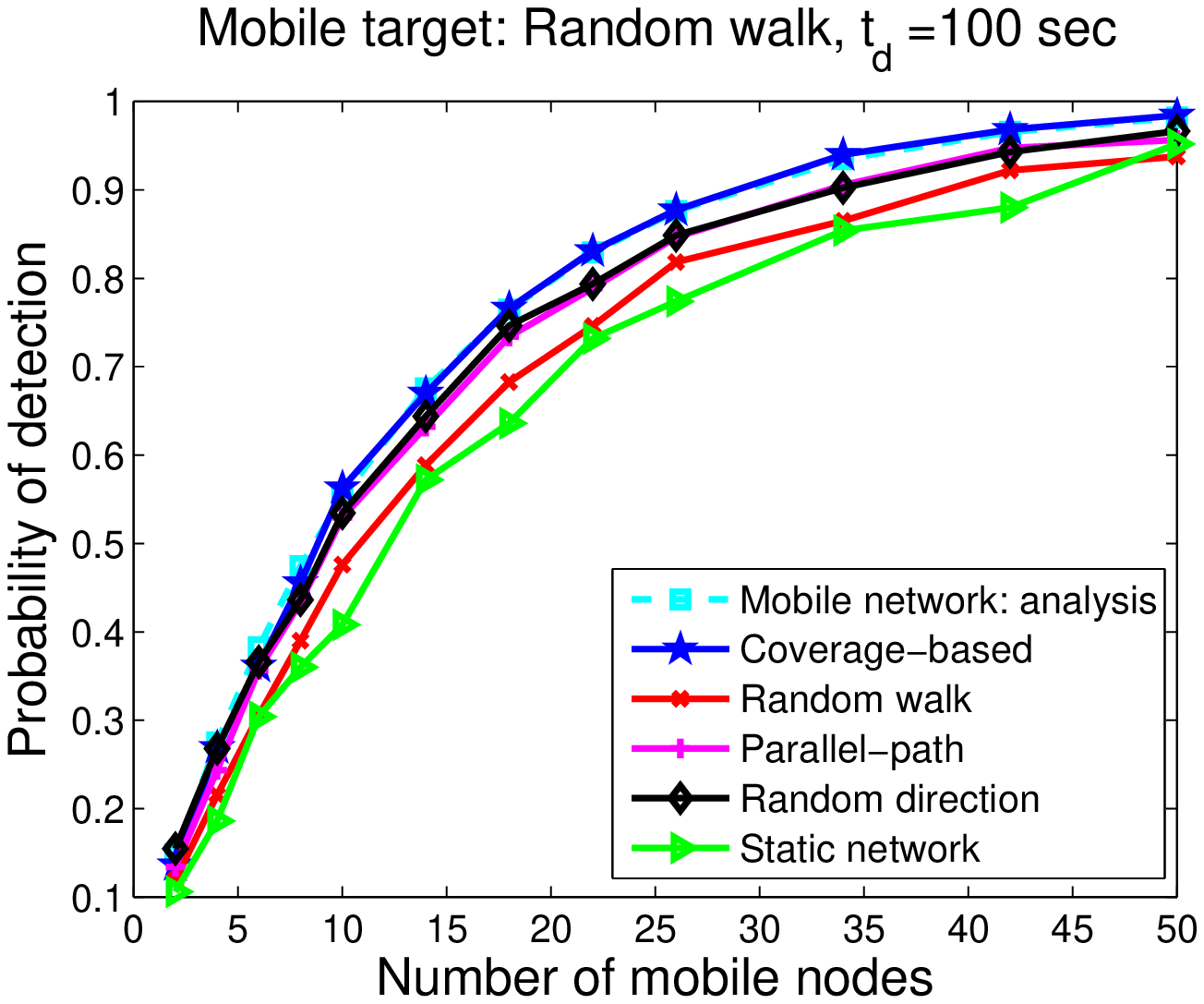,width=0.24\textwidth}}
\subfigure[]
{\psfig{figure=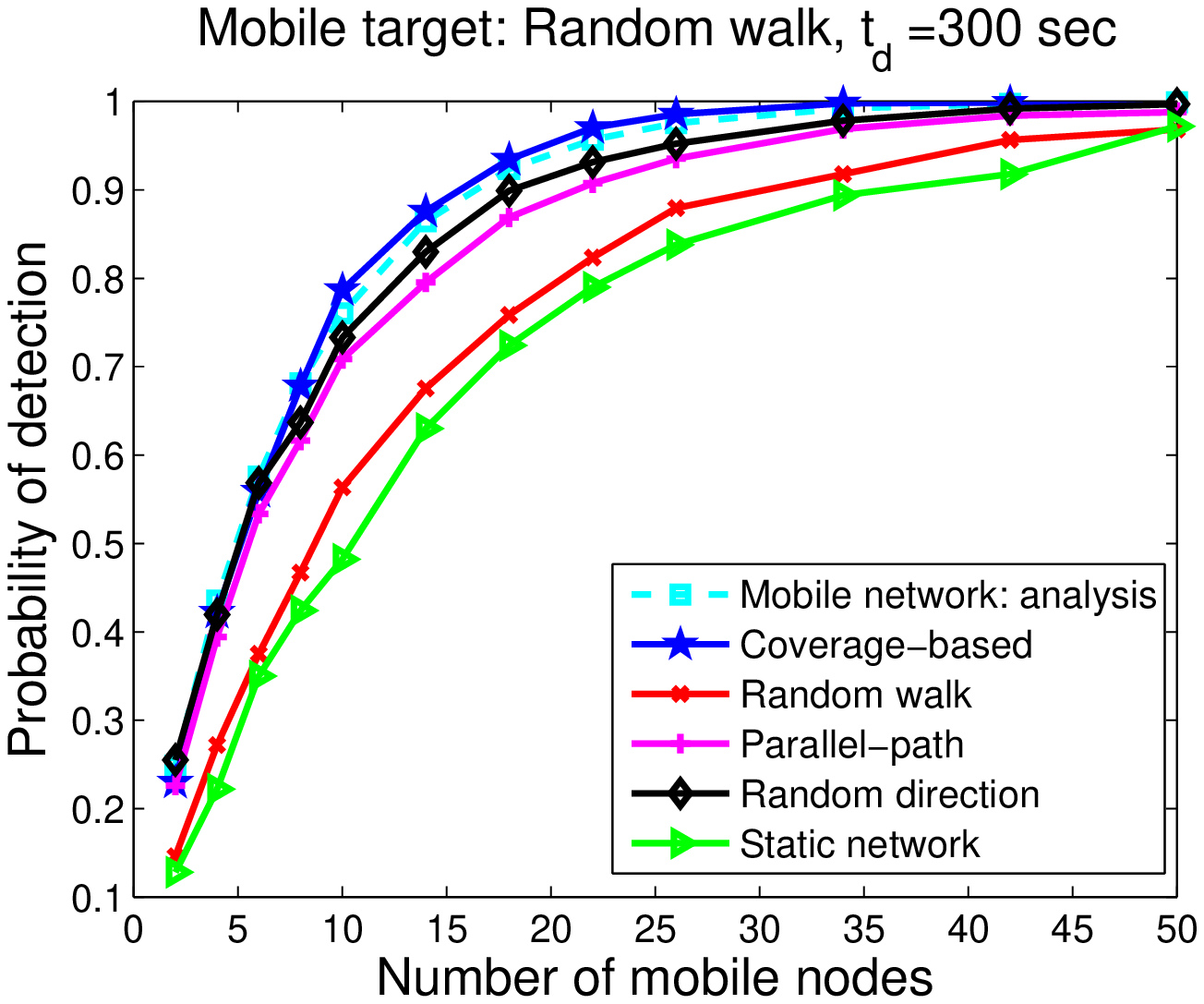,width=0.24\textwidth}}\\
\subfigure[]
{\psfig{figure=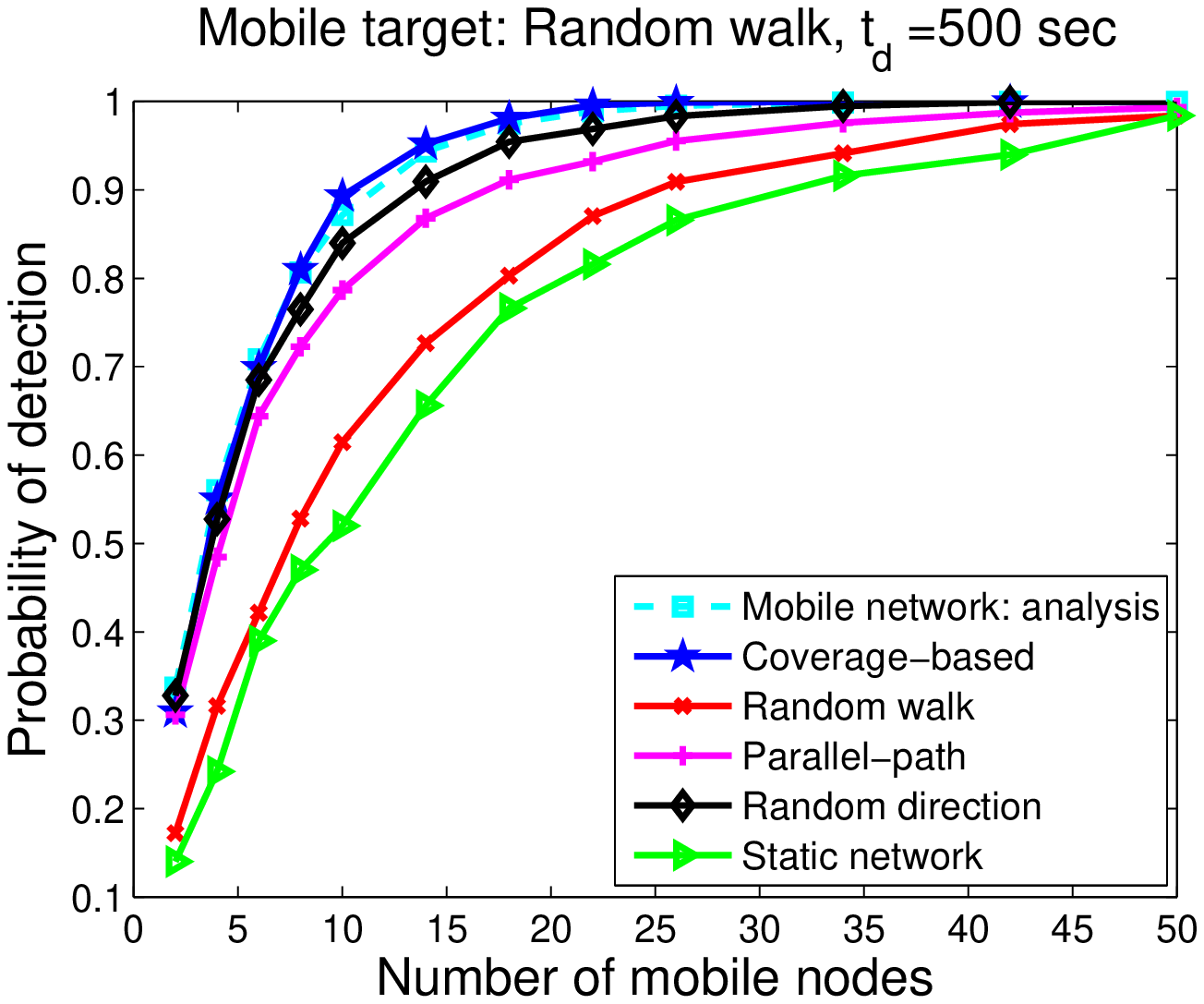,width=0.24\textwidth}}
\subfigure[]
{\psfig{figure=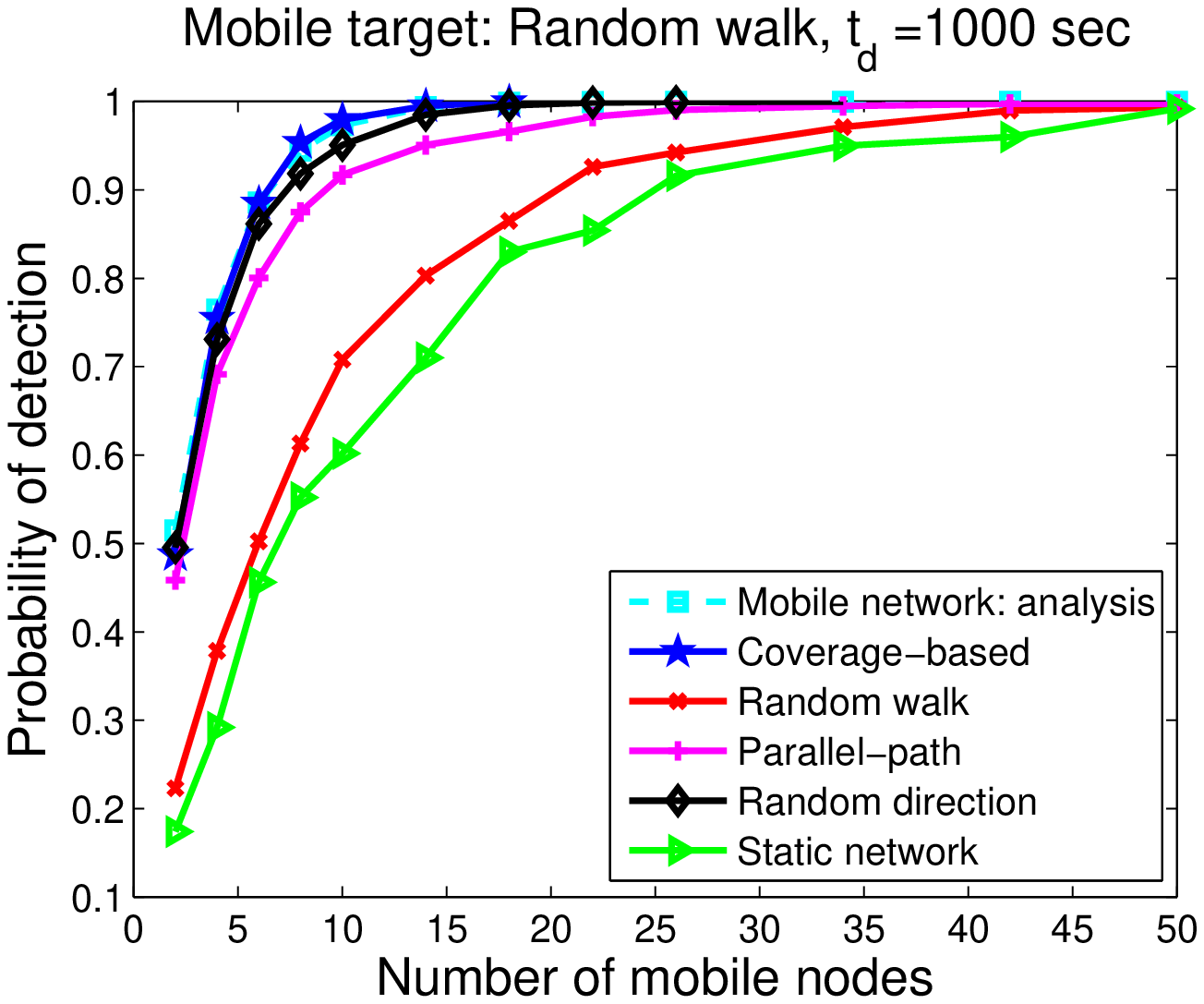,width=0.24\textwidth}}
\caption{Probability of detection versus number of nodes, when (a) $t_d = 100$s, (b) $t_d = 300$s, (c) $t_d = 500$s, (d) $t_d = 1000$s.}
\label{Pd_vs_Nm_RW}
\end{figure}

\begin{figure}[!htb]
\centering
\subfigure[]
{\psfig{figure=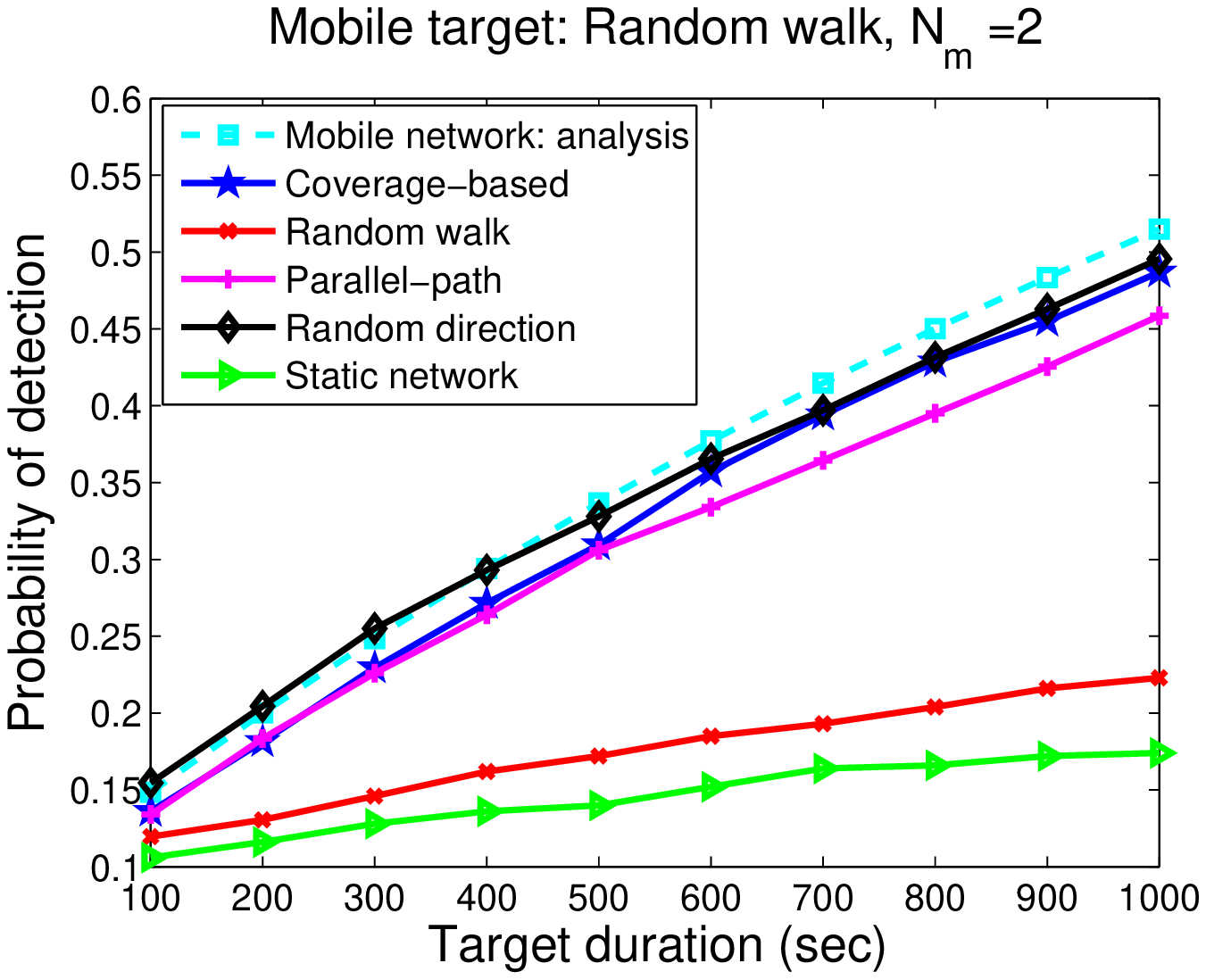,width=0.24\textwidth}}
\subfigure[]
{\psfig{figure=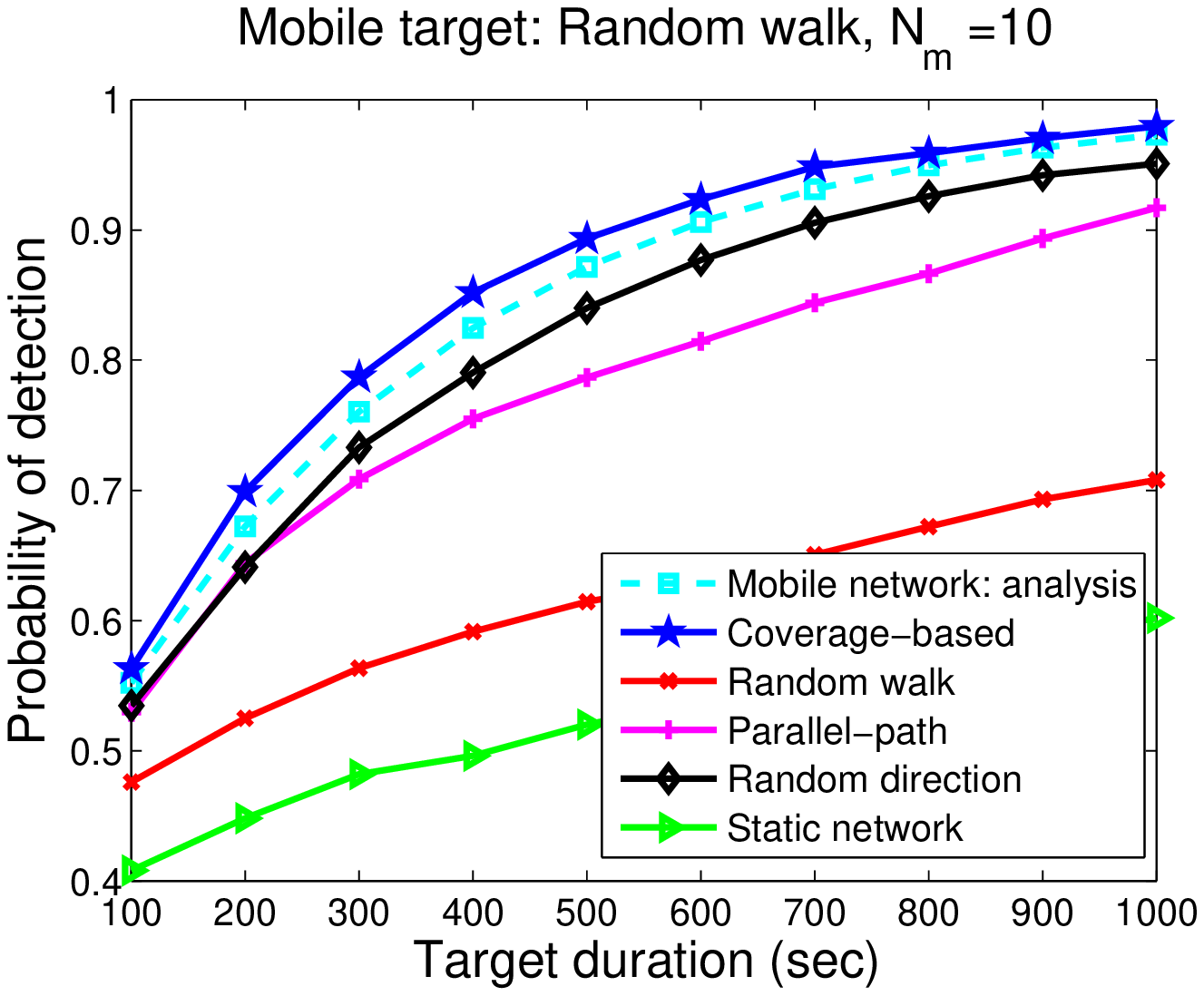,width=0.24\textwidth}}\\
\subfigure[]
{\psfig{figure=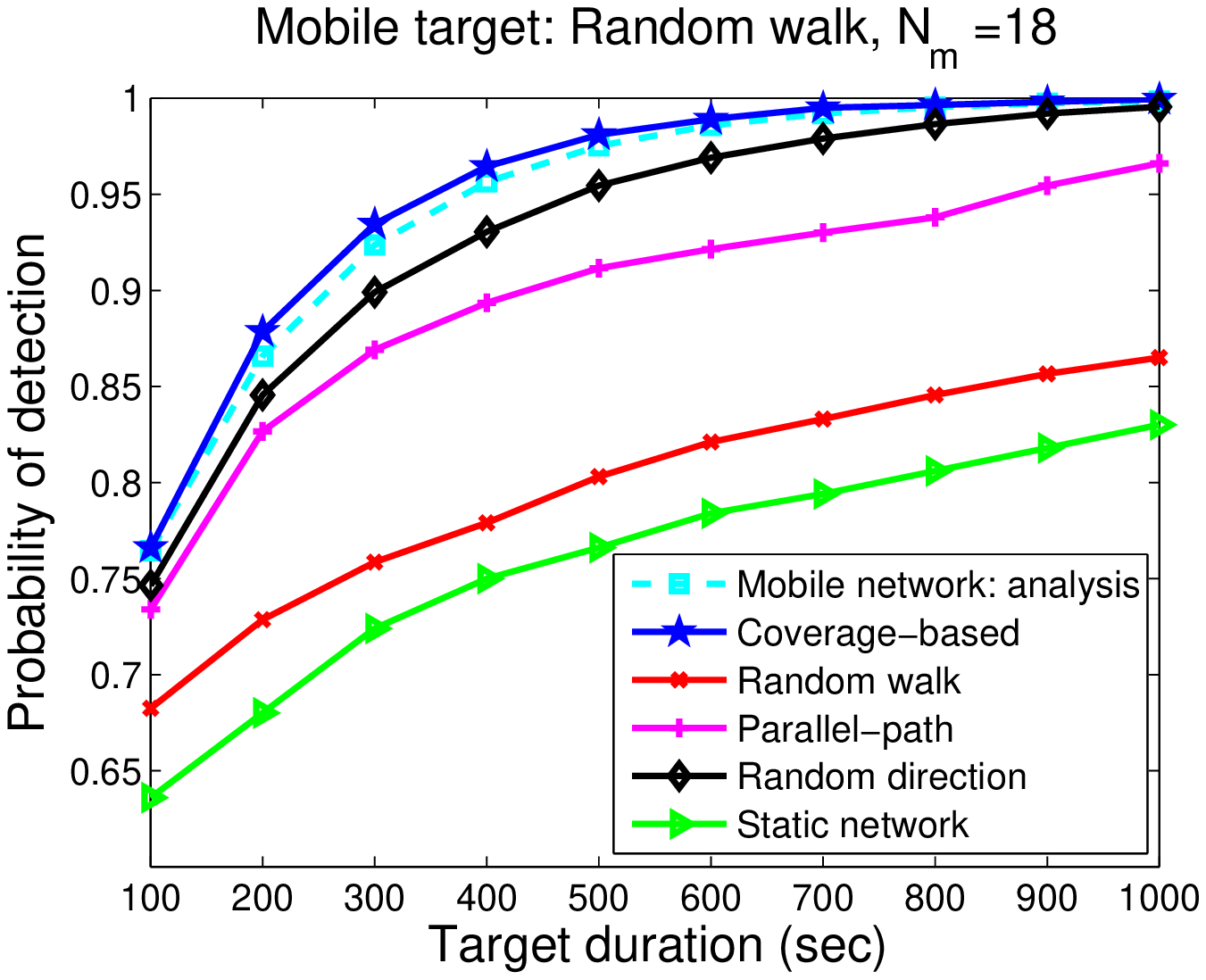,width=0.24\textwidth}}
\subfigure[]
{\psfig{figure=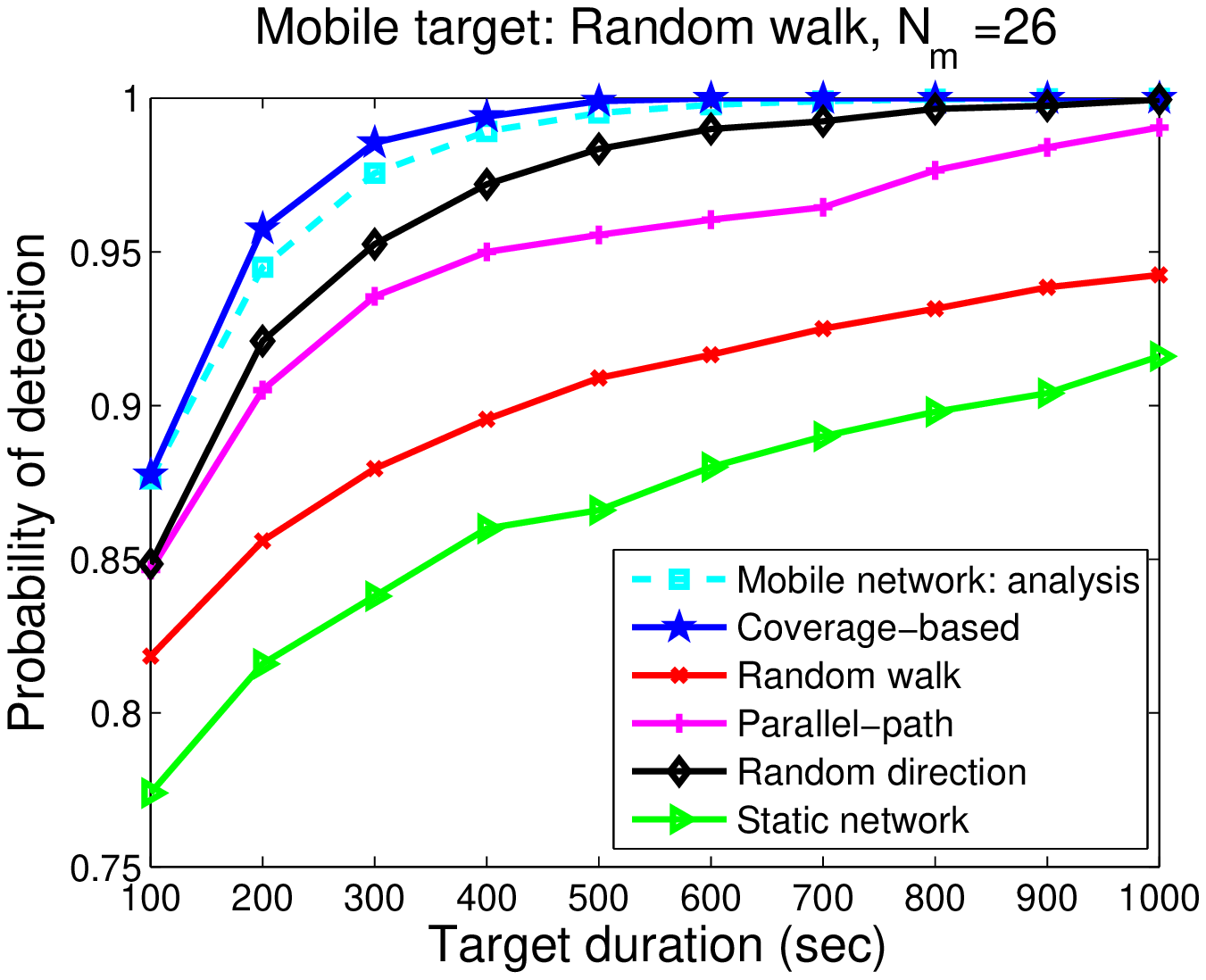,width=0.24\textwidth}}
\caption{{\small Probability of detection versus target duration, when (a) $N_m = 2$, (b) $N_m = 10$, (c) $N_m = 18$, and (d) $N_m = 26$.}}
\label{Pd_vs_td_RW}
\end{figure}

Fig.'s~\ref{Pd_vs_Nm_RW} and  \ref{Pd_vs_td_RW} present the detection probability performance of the mobile and static networks versus $N_m$ and $t_d$, respectively. The analytical results shown are obtained using Eq.~(\ref{Pcm_anl}). While this equation is derived for a stationary target, the simulation results for the coverage-based mobility model result in an excellent match to the analysis and provide an approximate estimate for the performance of the proposed model. Observe from Fig.~\ref{Pd_vs_Nm_RW} that static network performs as good as the mobile network with sensor nodes following a random walk, diminishing the benefit of mobility. However, the other mobility models outperform the static network as expected. As the number of mobile nodes is increased benefit of coverage-based model becomes more profound as shown in Fig.~\ref{Pd_vs_td_RW}.   

\begin{figure}[!htb]
\centering
\subfigure[]
{\psfig{figure=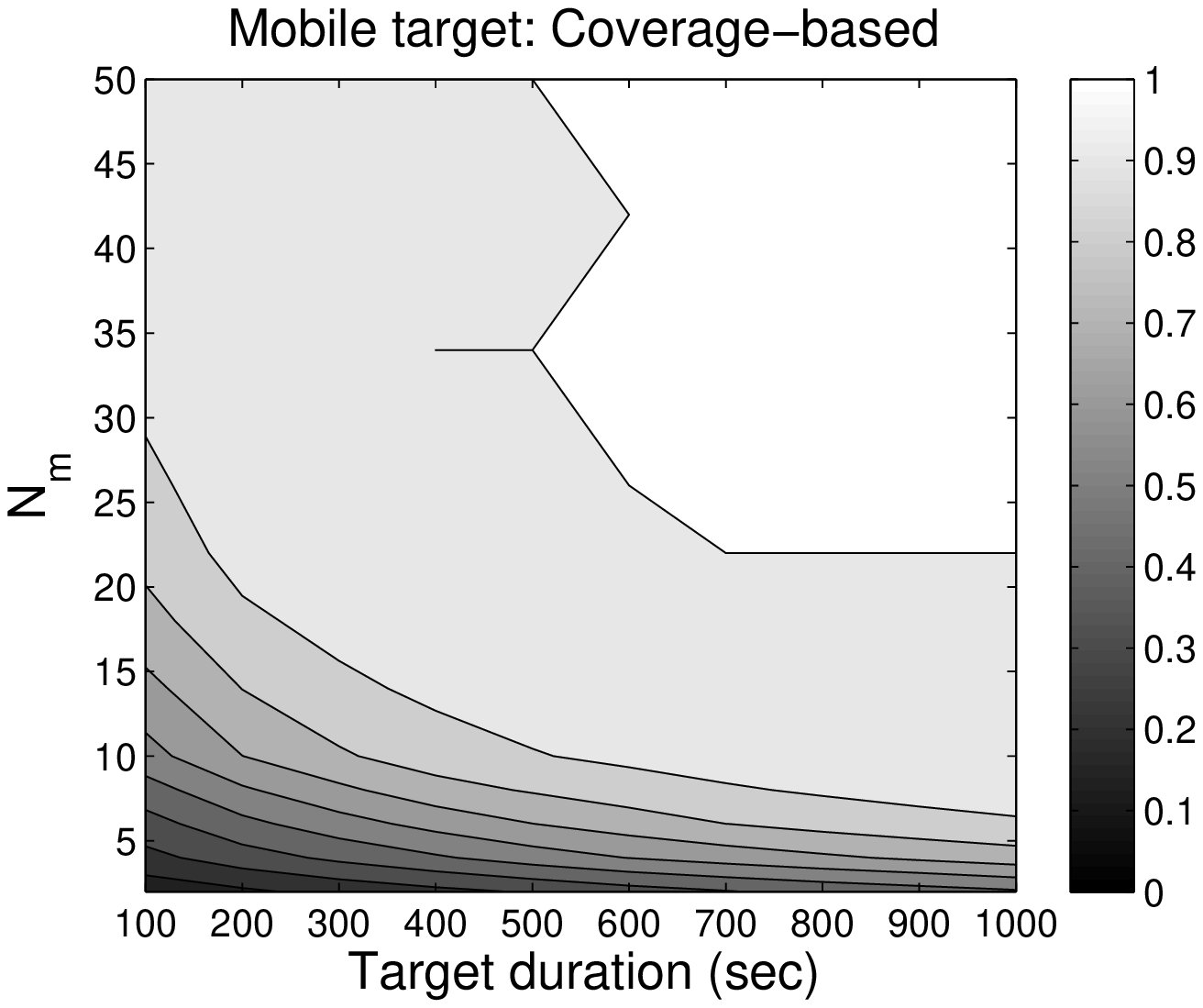,width=0.24\textwidth}}
\subfigure[]
{\psfig{figure=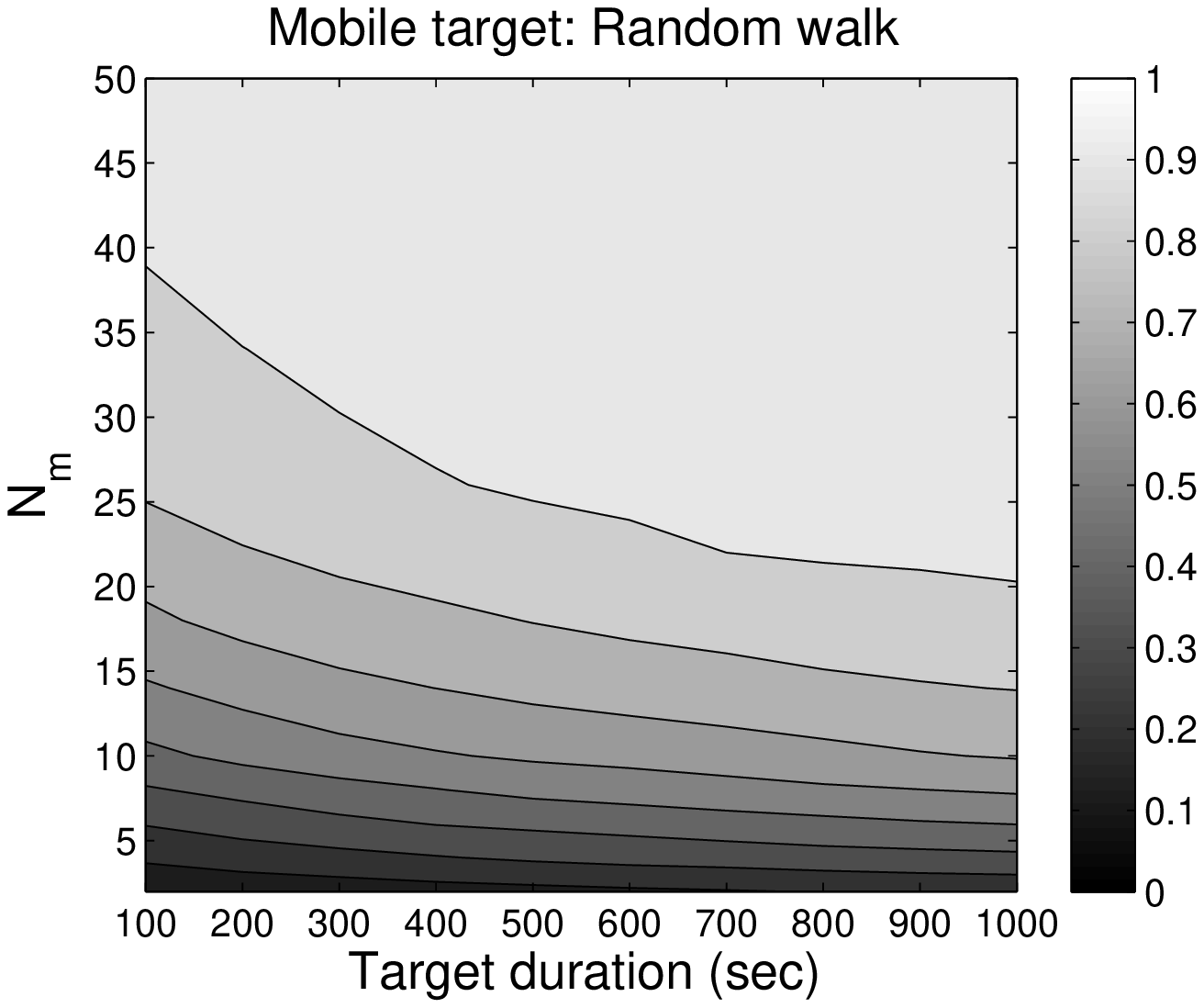,width=0.24\textwidth}}\\
\subfigure[]
{\psfig{figure=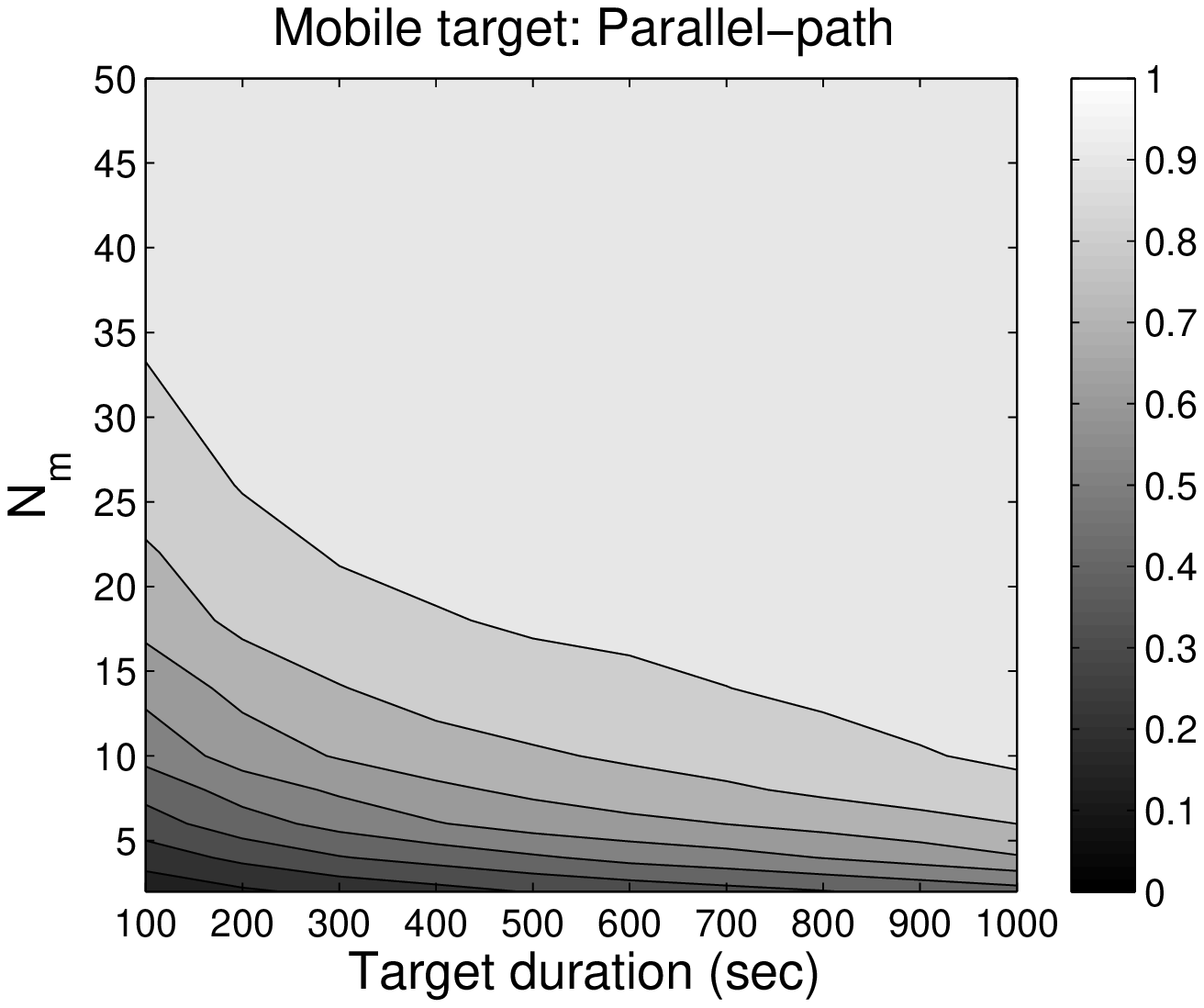,width=0.24\textwidth}}
\subfigure[]
{\psfig{figure=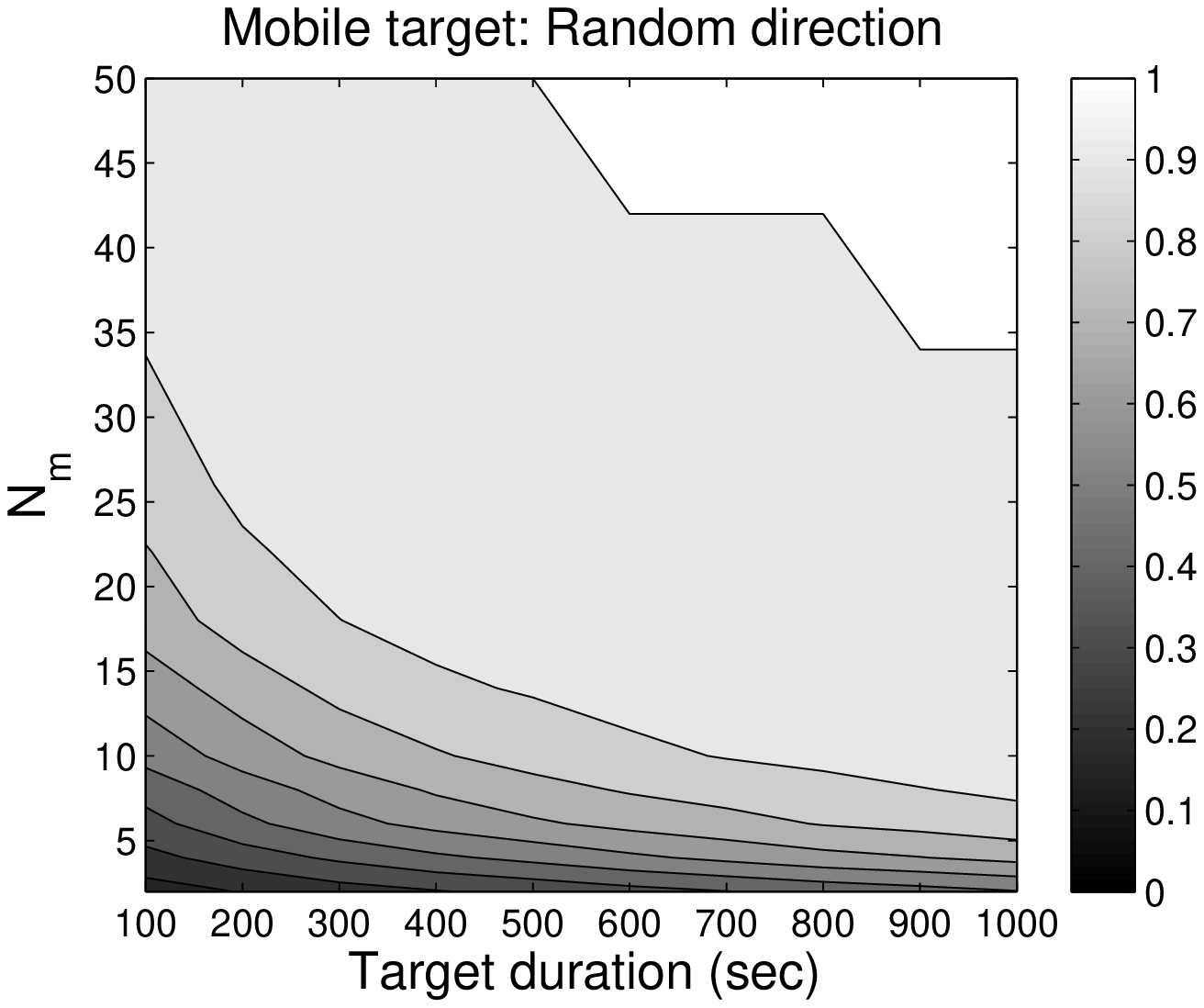,width=0.24\textwidth}}
\caption{Contour plots for probability of detection versus $N_m$ and $t_d$, for (a) Coverage-based, (b) Random walk, (c) Parallel-path, and (d) Random direction mobility models, when the target is mobile.}
\label{Pd_vs_Nm_td_RW}
\end{figure}

Finally, Fig.~\ref{Pd_vs_Nm_td_RW} shows the contour plot for the detection probability versus $N_m$ and $t_d$. Similar, to the stationary target scenario coverage-based model can achieve 100\% detection for a wider range of parameters. Different than the static case for the parameter values under investigation detection probabilities for all cases are higher as expected. 

\section{Conclusions}
In this work, a local, cooperative, coverage-based mobility model is proposed to improve stationary or mobile event coverage in mobile sensor networks. The proposed model uses local topology information and no application specific details are considered. The performance of the proposed model is compared with legacy mobility models in terms of target detection probabilities and tracking (monitoring) efficiencies. The results show that if the target detection probability requirements are highly stringent (i.e., the desired detection probability is close to 1), the benefit of the proposed model in terms of required mobile nodes become more significant. The results also illustrate the benefit of a mobile sensor network over a static network both in case of mobile and stationary targets. 
\bibliographystyle{ieee}

\bibliographystyle{latex8}

\end{document}